\documentclass[pra,a4paper,twocolumn,superscriptaddress,aps,notitlepage,10pt,longbibliography]{revtex4-1}

\usepackage{natbib}

\usepackage{times}
\usepackage{amssymb}
\usepackage{amsmath}
\usepackage{bm}
\usepackage{color}
\usepackage[colorlinks=true,urlcolor=blue,breaklinks,citecolor=blue]{hyperref}
\usepackage{graphicx}
\usepackage{braket}
\usepackage{subfigure}
\usepackage{xcolor}
\usepackage{float}
\usepackage{bbold} 

\definecolor{fuchsia}{rgb}{1.0, 0.0, 1.0}

\graphicspath{{.}{Images_Paper/}}

\begin{document}

\title{Characterizing the loss landscape of variational quantum circuits}

\author{Patrick Huembeli}
\affiliation{ICFO-Institut  de  Ciencies  Fotoniques,  The  Barcelona  Institute  of Science  and  Technology, 08860  Castelldefels  (Barcelona),  Spain}
\author{Alexandre Dauphin}
\affiliation{ICFO-Institut  de  Ciencies  Fotoniques,  The  Barcelona  Institute  of Science  and  Technology, 08860  Castelldefels  (Barcelona),  Spain}

\begin{abstract}
Machine learning techniques enhanced by noisy intermediate-scale quantum (NISQ) devices and especially variational quantum circuits (VQC) have recently attracted much interest and have already been benchmarked for certain problems.
Inspired by classical deep learning, VQCs are trained by gradient descent methods which allow for efficient training over big parameter spaces. For NISQ sized circuits, such methods show good convergence. There are however still many open questions related to the convergence of the loss function and to the trainability of these circuits in situations of vanishing gradients. Furthermore, it is not clear how ``good''
the minima are in terms of generalization and stability against perturbations of the data and there is, therefore, a need for tools to quantitatively study the convergence of the VQCs.
In this work, we introduce a way to compute the Hessian of the loss function of VQCs and show how to characterize the loss landscape with it. The eigenvalues of the Hessian give information on the local curvature and we discuss how this information can be interpreted and compared to classical neural networks. We benchmark our results on several examples, starting with a simple analytic toy model to provide some intuition about the behavior of the Hessian, then going to bigger circuits, and also train VQCs on data. Finally, we show how the Hessian can be used to adjust the learning rate for faster convergence during the training of variational circuits.
\end{abstract}

\maketitle

\section{Introduction}

Noisy intermediate-scale quantum (NISQ) devices~\cite{preskillQuantumComputingNISQ2018} have drawn a lot of attention in the last years and one of the most promising applications of NISQ devices are variational quantum circuits (VQCs). These quantum circuits (QC) are constructed with parametrized gates to minimize a given observable (or measurement) and can be trained with gradient based methods~\cite{stokesQuantumNaturalGradient2019a}.  There are now two main avenues for the use of VQCs. The first one is the use of VQCs as a state ansatz with the purpose of finding the variational parameters that minimize a given observable, such as the energy of a physical system for the Variational Quantum Eigensolvers (VQE)~\cite{mccleanTheoryVariationalHybrid2016, peruzzoVariationalEigenvalueSolver2014} or Quantum approximate optimization algorithms (QAOA)~\cite{farhiQuantumApproximateOptimization2014}. Such circuits allow one to study for example chemistry problems~\cite{caoQuantumChemistryAge2019} as well as classical optimization problems, where optimization tasks can be mapped to spin problems~\cite{farhiQuantumApproximateOptimization2014}. The second main application for VQCs encompasses data driven quantum machine learning (QML) tasks where VQCs are mainly treated as black boxes and their purpose is to map classical input data onto a measurement that serves as a label~\cite{farhiClassificationQuantumNeural2018,perez-salinasDataReuploadingUniversal2020, tacchinoArtificialNeuronImplemented2019a}. The emphasis in this application is to predict the right label for different classical inputs and the quantum states generated by VQCs are high dimensional representations of the classical data. These architectures of VQCs have shown to be universal approximators~\cite{perez-salinasDataReuploadingUniversal2020} and are therefore well suited to do machine learning. We will refer to this second application as Quantum Neural Network (QNN). Both of these applications have a classical Neural Network (NN) analogue. Classical NN architectures have been used as variational ansatz to represent quantum states ~\cite{carleoSolvingQuantumManyBody2017, hibat-allahRecurrentNeuralNetwork2020} and, like classical NNs, QNNs can be trained on data and do classification or even work as generative models~\cite{zoufalQuantumGenerativeAdversarial2019}. 

Recent works in VQCs were mainly devoted to the exploration of new applications. Nevertheless, the benchmark of the performance and advantage of such circuits compared to classical NNs has not yet received the deserved attention. In this work, we focus on the loss landscape of VQCs and characterize it with the help of the Hessian of the loss function. The loss landscapes of classical NNs trained on data have been extensively studied and we here compare the loss landscape of data driven classical NNs and QNNs. For data driven classical NN tasks, the study of the loss is almost as old as NNs themselves: it started with the investigation of spin glass models for a better understanding of Hopfield networks~\cite{hopfieldNeuralNetworksPhysical1982} and is still a very active research area. For example, numerical studies of over-parametrized classical NNs suggest that high error local minima traps do not appear~\cite{ballardEnergyLandscapesMachine2017, sagunExplorationsHighDimensional2015} in these high dimensional optimization landscapes. Empirical evidence even suggests that low loss basins in NNs are connected and there are no barriers separating them from each other \cite{draxlerEssentiallyNoBarriers2019}. Furthermore, different training algorithms find comparable solutions with similar training accuracies~\cite{sagunEmpiricalAnalysisHessian2018}. These findings are all in contradiction with the original idea of a spin-glass-like loss landscape, where many low energy solutions exist and one can get trapped in spurious minima. Supervised NN setups show good performance in generalizing tasks: once trained on data, they are able to predict labels of unknown test data. It is, until today, still the subject of debates what kind of topology the loss minimum of a well trained NNs should have.  
In Ref.~\cite{keskarLargeBatchTrainingDeep2017}, the authors showed that even though the training is independent of the gradient descent method, the generalization of the NNs depends on many hyper parameters, such as the batch size during the training: Smaller batch sizes tend to generalize better leading to wider minima basins of the loss function with many zeros eigenvalues of the Hessian $\lambda_i = 0$ and almost none of them negative $\lambda_i <0$. These findings support the idea that wider minima generalize better, whereas this question is not yet finally answered and is still subject to today's research in classical ML. The nature of the non-convexity of classical NNs is still in the focus of ongoing research~\cite{alainNegativeEigenvaluesHessian2019}.

A new phenomenon and a major short coming of VQCs compared to classical NNs is the occurrence of Barren Plateaus~\cite{mccleanBarrenPlateausQuantum2018a, cerezoCostFunctionDependentBarrenPlateaus2020, grantInitializationStrategyAddressing2019}, which are characterized by flat plateaus of the Loss function and exponentially vanishing gradients already for NISQ-device-size quantum circuits. In particular, the authors of Ref.~\cite{cerezoCostFunctionDependentBarrenPlateaus2020} showed that the appearance of the Barren plateaus depends on the choice of the cost function and that for local cost functions, gradient decays only with a power law. The authors also claimed that the loss landscape of certain problems resembles  a narrow gorge, which somehow contrasts with the idea of wide basins for good generalization in classical NNs. In Ref.~\cite{grantInitializationStrategyAddressing2019}, the authors showed that one can avoid to initialize the variational parameters of the circuit in a Barren Plateau. In so called hybrid-quantum-classical training settings, standard gradient methods like Stochastic gradient descent (SGD), Adam or also the quantum natural gradient (QNG) have been used to train VQCs. QNG showed good convergence~\cite{stokesQuantumNaturalGradient2019a} and also helped to avoid local minima~\cite{wierichsAvoidingLocalMinima2020b}, which is in contrast to the aforementioned independence of the training method for classical NNs. In Ref.~\cite{perez-salinasDataReuploadingUniversal2020}, the authors used a Hessian based optimization method and obtain faster convergence for certain data driven problems. Later in this work, we will provide a possible explanation for this phenomenon.

Generally the loss landscape of VQCs is not yet well understood and, as for classical NNs, a better understanding might lead to the improvement of optimization algorithms and show the limitations of certain circuit designs. In this work we show how the loss landscape of different VQCs can be studied via the eigenvalues of the Hessian. 

The paper is structured as follows: In SubSecs.~\ref{sec:Hessian_and_curvature} and \ref{sec:LL_of_cl_NN}, we start with a brief introduction of the Hessian of NN losses, the loss landscape of classical NNs and the distribution of the eigenvalues of the Hessian. In SubSec.~\ref{sec:LL_of_VQCs}, we discuss the loss landscapes of VQCs. In Sec.~\ref{sec:loss_landscape}, we study an analytical example of a VQC and compute the Hessian and its eigenvalues. We show how to calculate the Hessian on an actual quantum hardware in Sec.~\ref{sec:Hessian_of_QC}, apply it to a general example in Sec.~\ref{Sec:General_Circuit} and study it numerically. In Sec.~\ref{sec:QNN_data}, we study a VQC trained on classical data acting as a classifier. Finally in Sec.~\ref{sec:Barren_Plateau}, we show how the Hessian can help to escape from very flat barren-plateau-like regions in the landscape.

\section{Charaterization of the loss landscape with the Hessian}
\subsection{Hessian and curvature}
\label{sec:Hessian_and_curvature}

The Hessian of a function is a well suited mathematical object to study the local curvature of a function. Given a function $f(\bm{\theta}): \mathbb{R}^N \rightarrow \mathbb{R}$, the Hessian is defined as the square matrix of the second derivatives of this function $H_{ij} = \partial_{\theta_i} \partial_{\theta_j} f(\bm{\theta})$, with $\bm{\theta} = (\theta_1, \theta_2, \dots, \theta_N)$. For functions satisfying the Schwartz theorem, the Hessian is a symmetric matrix and its eigenvalues $\lambda_i$ are real and can be ordered $\lambda_1 \leq \lambda_2 \leq \dots \leq \lambda_N$. We denote by $\bf{v}_i$ the eigenvector associated with the eigenvalue $\lambda_i$ and refer to the pair $(\lambda_i, \bf{v}_i)$ as the i-th eigenpair. 

The Hessian $H \rvert_{\bm{\theta}}$  evaluated at a certain point $\bm{\theta}$ gives a second order approximation of the function $f(\bm{\theta})$ within Taylor's expansion and its spectrum  allows one to extract information about the local curvature of $f(\bm{\theta})$: A positive eigenvalue $\lambda_i > 0$ of an eigenpair indicates a locally positive curvature in the direction of $\bm{v}_i$ and, therefore, an increase of $f(\bm{x})$ if one moves along the direction $\bm{\theta} + \epsilon \mathbf{v}_i$, for a small perturbation~$\epsilon$. Analogously, negative eigenvalues indicate a locally negative curvature and $\lambda_i = 0$ indicate flat directions of the function $f(\bm{\theta})$ and, consequently, zero curvature. Accordingly, if the function is at an extremum $\nabla f(\bm{\theta}) = 0$ and all eigenvalues are positive ( resp. negative), the point is a local minimum (maximum). If some eigenvalues are positive and some are negative the extremum is a saddle point. 

\subsection{Loss landscape of neural networks: a brief review}
\label{sec:LL_of_cl_NN}

The training of a NN in a supervised machine learning setting consists in minimizing the empirical risk $\mathcal{L} = \sum_i^N l(f(\vec{x}_i, \bm \theta), y_i)$ over a dataset  $\mathcal{D}= \{ (\vec{x}_i, y_i) \}_i^N$, where $x_i$ is a training point with its corresponding label $y_i$. 
$f(\vec{x_i}, \bm \theta)$ is the NN prediction for the training point $x_i$ parameterized by the weights $\bm \theta$. The loss function $l(\cdot)$ measures the difference between the NN prediction and the label. Commonly used loss functions are the mean square error or the cross entropy~\cite{mehtaHighbiasLowvarianceIntroduction2019a}. 
The Hessian is defined as $H_{ij} = \partial_{\theta_i\theta_j} \mathcal{L}$ and has a wide range of applications in ML: It can be used to adapt gradient update to the current loss landscape in the so called "Newton" method~\cite{leOptimizationMethodsDeep}, for pruning~\cite{NIPS1989_250,hassibi1993second} or for interpretability purposes with the influence function~\cite{kohUnderstandingBlackboxPredictions2017b}. Furthermore, it can also be used to study the local curvature of the loss for a better understanding of the loss landscape and the convergence of NNs.
In an extensive empirical study, the authors of Ref.~\cite{sagunEmpiricalAnalysisHessian2018} charaterized the behaviour of the Hessian for over-parameterized NNs, where the number of weights $\bm \theta$ of the NN is much bigger than the number of training points $N$. They showed that, after training, the eigenvalues of the Hessian are distributed such that most eigenvalues are in a bulk close to zero and that there are some outliers away from zero. For a randomly initialized NNs, the outliers are both positive and negative and they are symmetrically distributed around zero. With the training progress, the amount of positive eigenvalues increases and it converges to a negligible amount of negative eigenvalues. For a trained NNs most eigenvalues are zero and a few are positive, which indicates a flat minimum of the loss landscape. These findings suggest that classical notions of basins of attraction may be misleading and the minima of over parameterized NNs are extremely flat pools with just a few steep directions. The authors also showed that the number of eigenvalues that are significantly larger than zero can be even smaller than the number of training points $N$ given that the training data instances of each class do not deviate significantly. They can be of the order of $k$, which is the number of classes that have to be learned by the NN.

\begin{figure}[t]
\centering
  \includegraphics[width=\linewidth]{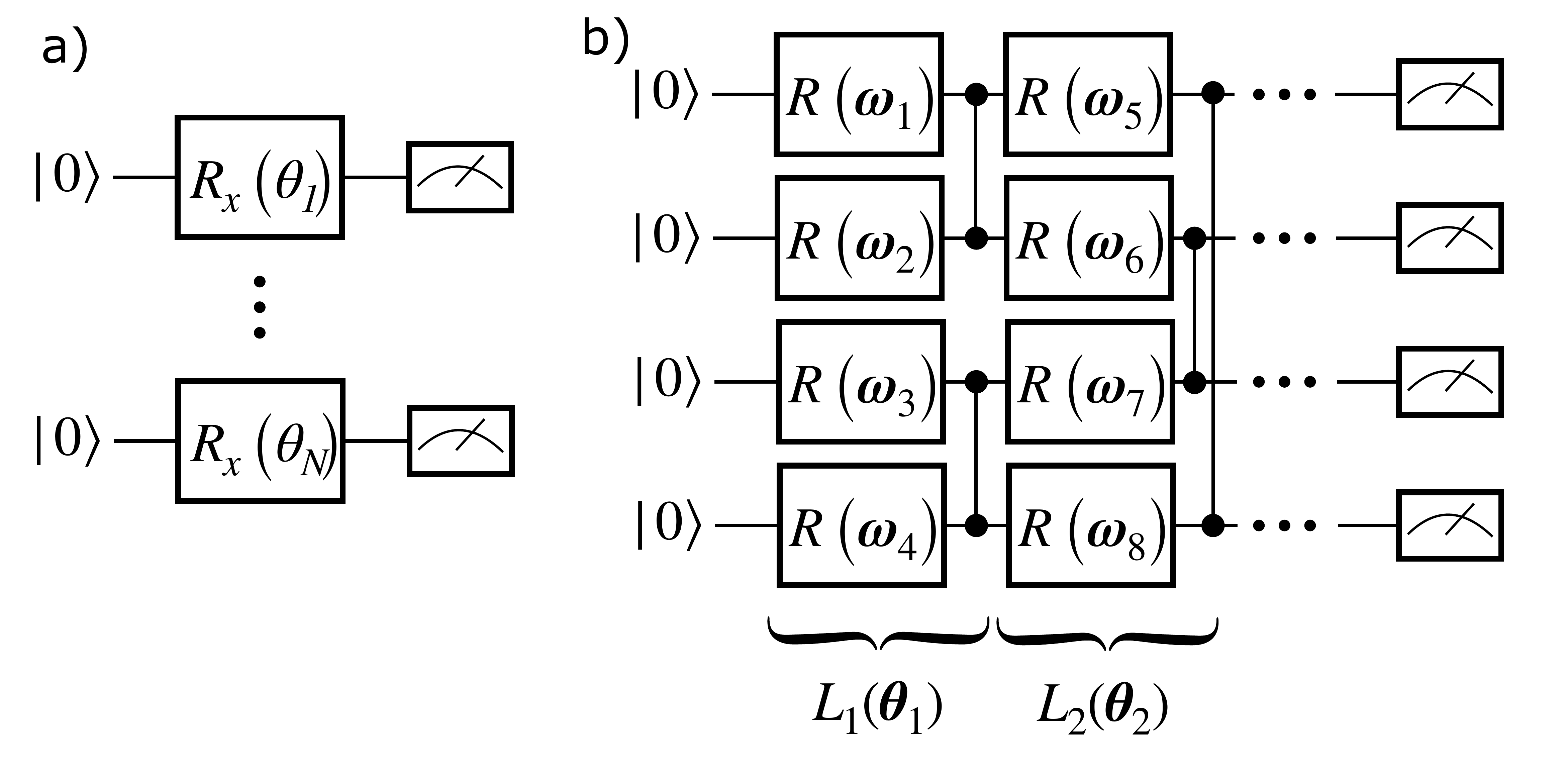}
  \caption{\textbf{Variational Circuits.} a) Toy model circuit with 1 qubit rotations around the x axis for $N$ qubits without entangling gates. b) A general 4 qubit circuit with 1 qubit rotations $R(\boldsymbol{\omega}_i) = R(\phi_1, \phi_2, \phi_3) $ followed by CZ entangling gates. We denote by $L_i$ is a combination of rotations and entangling gates.}
  \label{fig:Toy_and_General_Circuit}
\end{figure}

\subsection{Loss landscape of VQCs}
\label{sec:LL_of_VQCs}

For the study of loss landscapes of VQCs the literature, so far, is rather sparse and there are still many open questions. For pure quantum optimization tasks, such as QAOA, VQE or also state preparation, in general, Barren plateaus seem to be a major shortcoming and $\nabla f(\bm{\theta}) = 0$ might not indicate an extremum in the loss landscape but an absolutely flat region, where all eigenvalues of the Hessian are zero. In Ref.~\cite{cerezoCostFunctionDependentBarrenPlateaus2020}, the authors defined a Barren Plateau as a point in the parameter space, 
where the mean gradient $\braket{\nabla f(\bm{\theta}) }= 0$ and the variance $\text{Var} \left[ \nabla f(\bm{\theta}) \right]$ exponentially vanish with the number of qubits $N$ for global cost functions and any circuit depth $d$. 
They also showed that Barren plateaus can be avoided for local cost functions if the circuit depth is $d \in \mathcal{O}(\log(N))$. 
For data driven tasks, where VQCs are trained on classical data, which we will refer to as a Quantum Neural Network (QNN), the occurrence of Barren Plateaus has been observed in some specific architecutres~\cite{sharma2020trainability}.
In section~\ref{sec:QNN_data}, we investigate a QNN which seems to not suffer from vanishing gradients. The latter might come from the fact that such architecture escapes the ones described in Ref.~\cite{sharma2020trainability}. The loss landscape for VQCs has so far been studied in the context of the Quantum Natural Gradient (QNG)~\cite{stokesQuantumNaturalGradient2019a, wierichsAvoidingLocalMinima2020b} where it has been shown that including the local curvature of the quantum state can improve training convergence and also help to avoid local minima. Furthermore, in VQC settings, where the aim is to find a target state $\ket{\psi_T}$, the question of how expressive a certain circuit ansatz is and whether the ansatz contains a solution to the problem becomes relevant~\cite{simExpressibilityEntanglingCapability2019}. Unfortunately, if one wants to find a target state with gradient based optimization methods, it will not be enough to know that the ansatz contains a solution. It might be impossible to reach it from a certain initialization of the parameterized circuit, because one gets stuck in local minima. This problem has been studied for classical NNs and empirical evidence suggests that for overparametrized NNs the local minima basins are connected~\cite{keskarLargeBatchTrainingDeep2017a} and therefore overparametrization is as crucial as the bare existence of the solution itself. It is still unclear if these notions of overparameterization also apply to VQCs. Hence, there is a need for a better understanding of the loss landscapes of general VQCs and we believe that the use of the Hessian is one possible tool to achieve this.

\section{Loss landscape of VQCs: an analytical example}
\label{sec:loss_landscape}
\begin{figure}[t]
  \centering
  \begin{minipage}[b]{0.23\textwidth}
    \includegraphics[width=\textwidth]{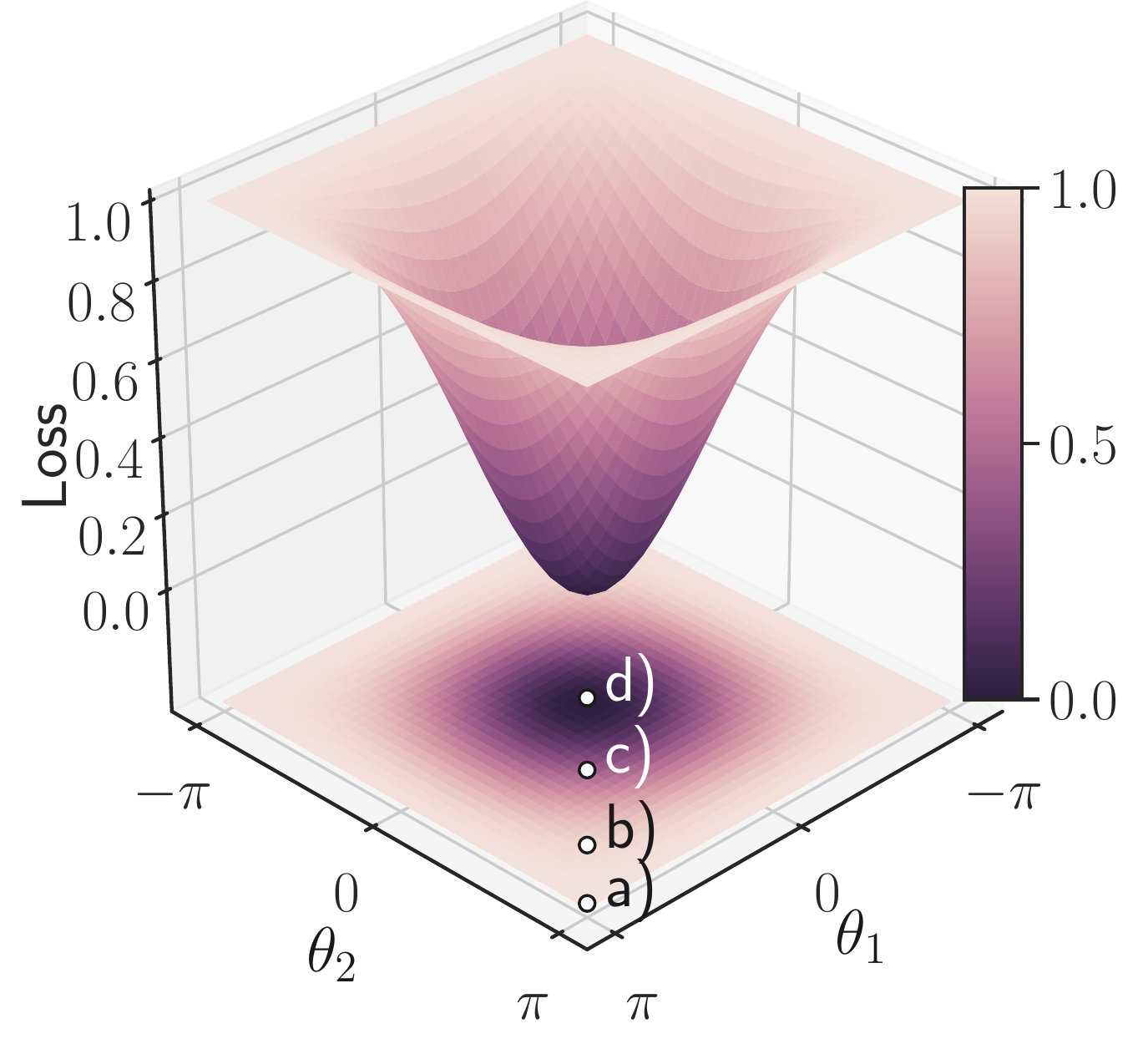}
  \end{minipage}
  \hfill
  \begin{minipage}[b]{0.23\textwidth}
    \includegraphics[width=\textwidth]{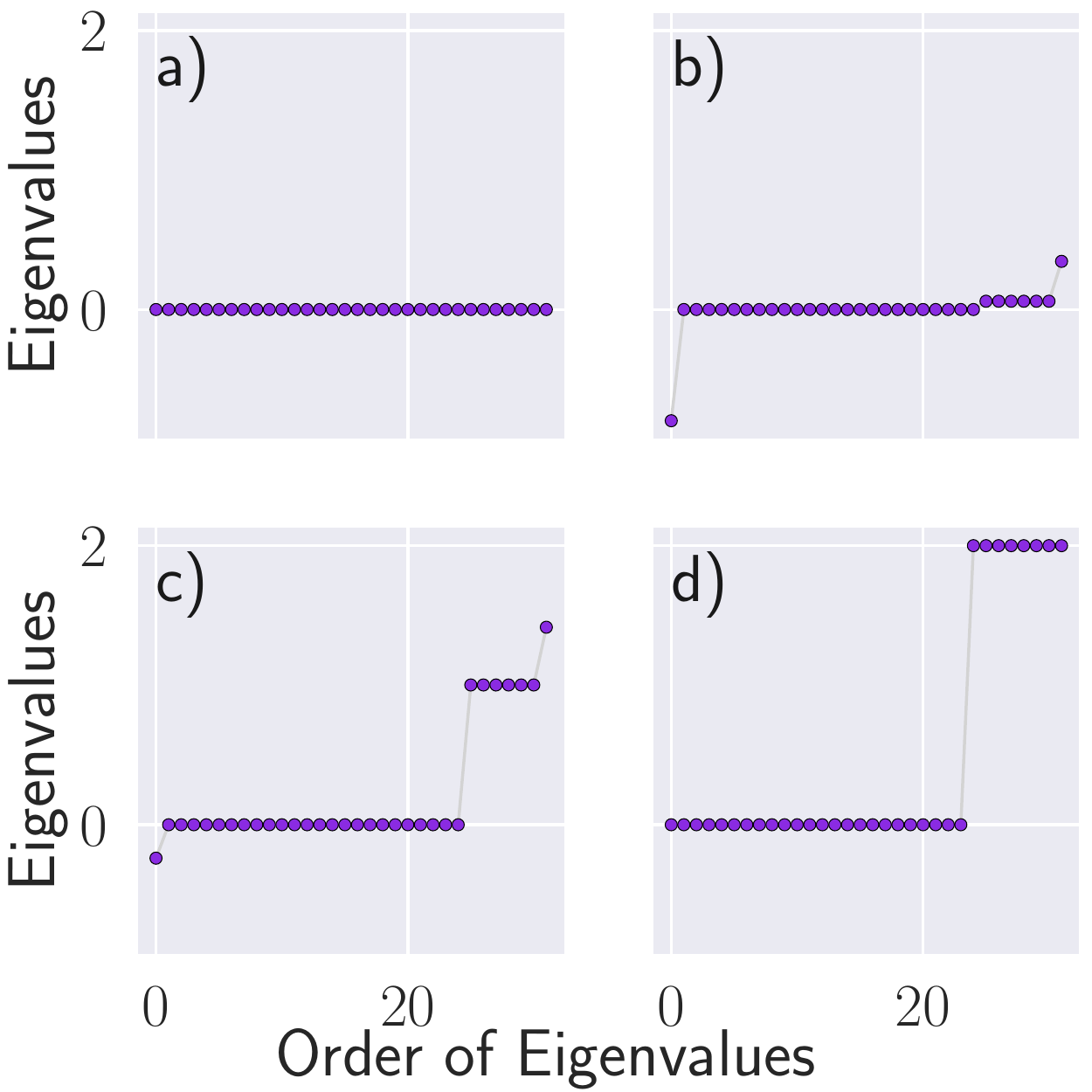}
  \end{minipage}
      \caption{\textbf{Loss landscape of the toy model with a global loss function.} The target state is chosen to be $\ket{\psi_T} = \ket{\mathbf{0}}$. We label in the contour plot the points for which we compute the Hessians and its eigenvalues. For fixed parameters $\theta_i=0$ with $i>2$ the loss function is not dependent on the number of qubits. Panel a) shows the vanishing eigenvalues which indicate the Barren Plateau described in Ref.~\cite{cerezoCostFunctionDependentBarrenPlateaus2020}.}
  \label{fig:Loss_landscape_toy_simple}
\end{figure}

In this section we analytically characterize the curvature of the loss landscape with the Hessian. We start with a simple circuit that can be solved analytically to get familiar with the concept of the Hessian of loss functions of VQCs. We choose a toy model circuit introduced in Ref.~\cite{cerezoCostFunctionDependentBarrenPlateaus2020} that presents Barren Plateaus. The toy model is a $N$ qubit VQC $V(\boldsymbol{\theta}) =  \otimes_i^N R_x(\theta_i)= \otimes_i^N \exp(-i \theta_i/2 \sigma_x)$, shown in Fig.~\ref{fig:Toy_and_General_Circuit}(a), with randomly initialized parameters $\boldsymbol{\theta}$ that generates the state 
\begin{equation}
    V(\boldsymbol{\theta}) \ket{\mathbf{0}} =  \sum_{k=0}^n (i)^k P \left( \prod_i^{n-k} \cos( \frac{\theta_i}{ 2}) \prod_j^k \sin( \frac{\theta_j}{ 2}) \ket{0}^{n-k} \ket{1}^{k} \right), 
    \label{Eq:General_Target}
\end{equation}
where $\ket{\mathbf{0}}\equiv \otimes_i^N  \ket{0}_i$ and $P(\cdot)$ stands for the sum of all possible permutations of the argument. For example
\begin{equation}
\begin{split}
&P\left(\cos(\theta_1) \sin(\theta_2) \ket{01}\right) \\
&= \cos(\theta_1) \sin(\theta_2) \ket{01} + \cos(\theta_2) \sin(\theta_1) \ket{10}.
\end{split}
\end{equation} 
The aim of the variational algorithm is to rotate the initial state $V(\boldsymbol{\theta}) \ket{\mathbf{0}}$ into a given target state $\ket{\psi_T}$ and to maximize the fidelity $\mathcal{F}$ with respect to $\ket{\psi_T}$. For the particular target state $\ket{\psi_T} = \ket{\mathbf{0}}$, all the $\sin( \cdot)$ terms in the variational state of Eq.~\eqref{Eq:General_Target} cancel and the fidelity reads~\cite{cerezoCostFunctionDependentBarrenPlateaus2020}
\begin{equation}
    \mathcal{F}= \vert\bra{\psi_T} V(\boldsymbol{\theta}) \ket{\boldsymbol{0}}\vert^2 = \prod_i^n \cos^2 \left( \frac{\theta_i}{2} \right).
    \label{Eq: Loss global Toy model}
\end{equation}
Therefore, we can translate this optimization problem into the minimization of the loss function $l = 1 - \mathcal{F}$, shown in Fig.~\ref{fig:Loss_landscape_toy_simple}, as a function of $\theta_1$ and $\theta_2$ and for $\theta_{i>2}=0$.

We now discuss how the Hessian can help to understand the loss landscape. 

First we initialize the parameters $\bm\theta$ randomly and we want to find the parameters to generate the target state. Figure~\ref{fig:Loss_landscape_toy_simple} depicts the eigenvalues of the Hessian for different values of $\theta_1$ and $\theta_2$. The points a) - d) in the optimization landscape show a possible trajectory of an optimizer. Point a) shows an initialization in a Barren Plateau, where all the eigenvalues of the Hessian are 0. Points b) and c) are non-extremal  points in this high dimensional optimization space where some of the eigenvalues are negative, some are positive and the bulk of them is zero. Point d) shows a well converged loss with no negative eigenvalues of the Hessian and zero gradient ensuring that it is indeed a local minimum. Since the loss $l=0$, we even know that it is a global minimum. We also observe many degenerate zero eigenvalues which implies that the minimum is not isolated. As it was the case for classical NNs, this is a consequence of the over parametrization of the VQC: many different linear combinations of angles lead to the same loss function.

For two free parameters $\theta_1$ and $\theta_2$ and the rest fixed, we find that the amplitudes of the eigenvalues of the Hessian do not depend on the system size.
This might seem to contradict the phenomenon of narrow gorge described in Ref.~\cite{cerezoCostFunctionDependentBarrenPlateaus2020}, where the authors describe that the valley in Fig.~\ref{fig:Loss_landscape_toy_simple} becomes narrower with an increasing number of qubits $N$. But in their work, they change half of the parameters collectively along one axis and the other half along the other axis. Therefore, it is still true that the VQC is more likely to be initialized in a flat region even with our description, because the parameter space becomes bigger and the product of sinusoidal functions of Eq.~\eqref{Eq:General_Target} becomes smaller for a larger number of randomly drawn parameters.

\begin{figure}[t]
  \centering
  \begin{minipage}[b]{0.23\textwidth}
    \includegraphics[width=\textwidth]{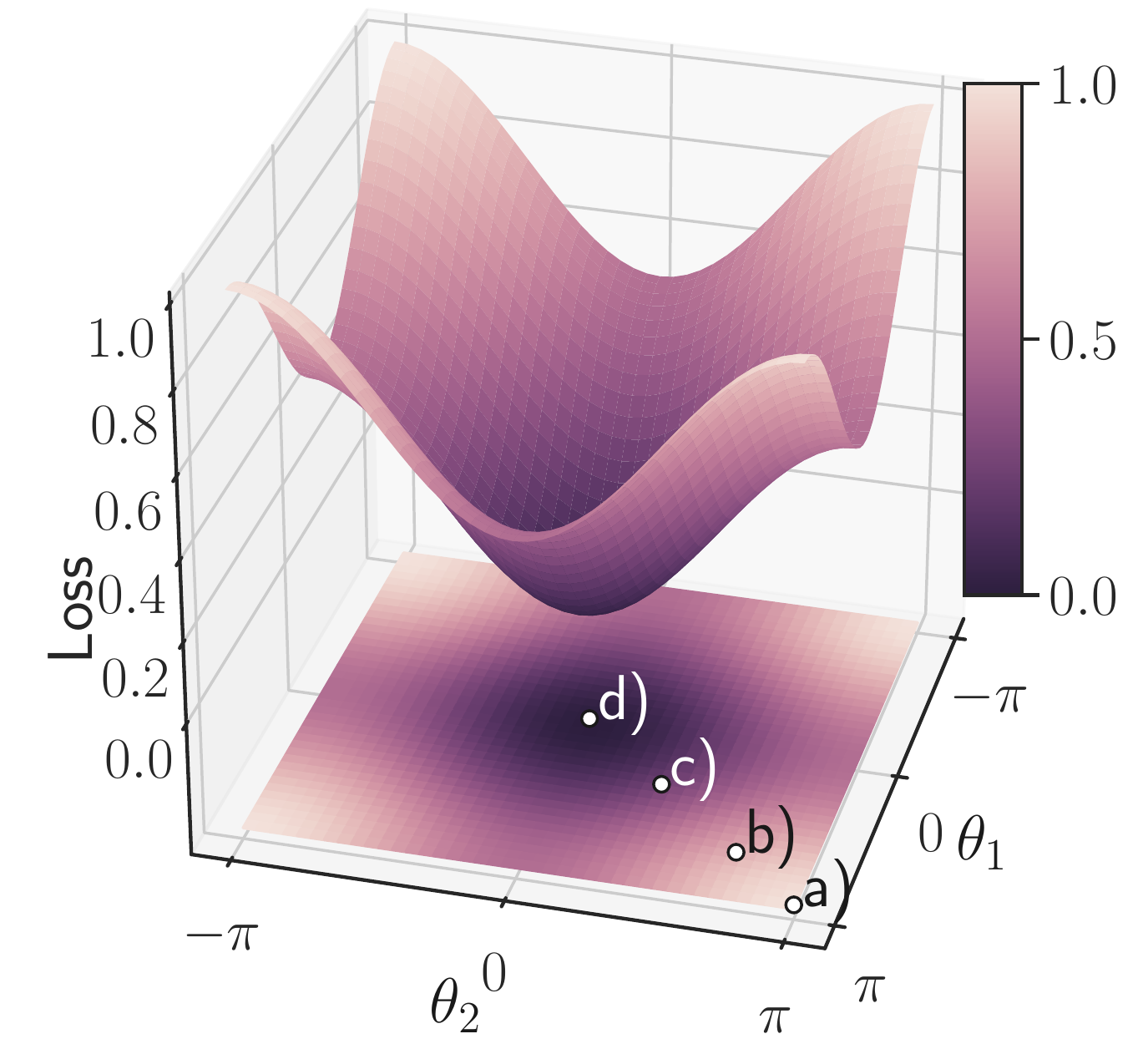}
  \end{minipage}
  \hfill
  \begin{minipage}[b]{0.23\textwidth}
    \includegraphics[width=\textwidth]{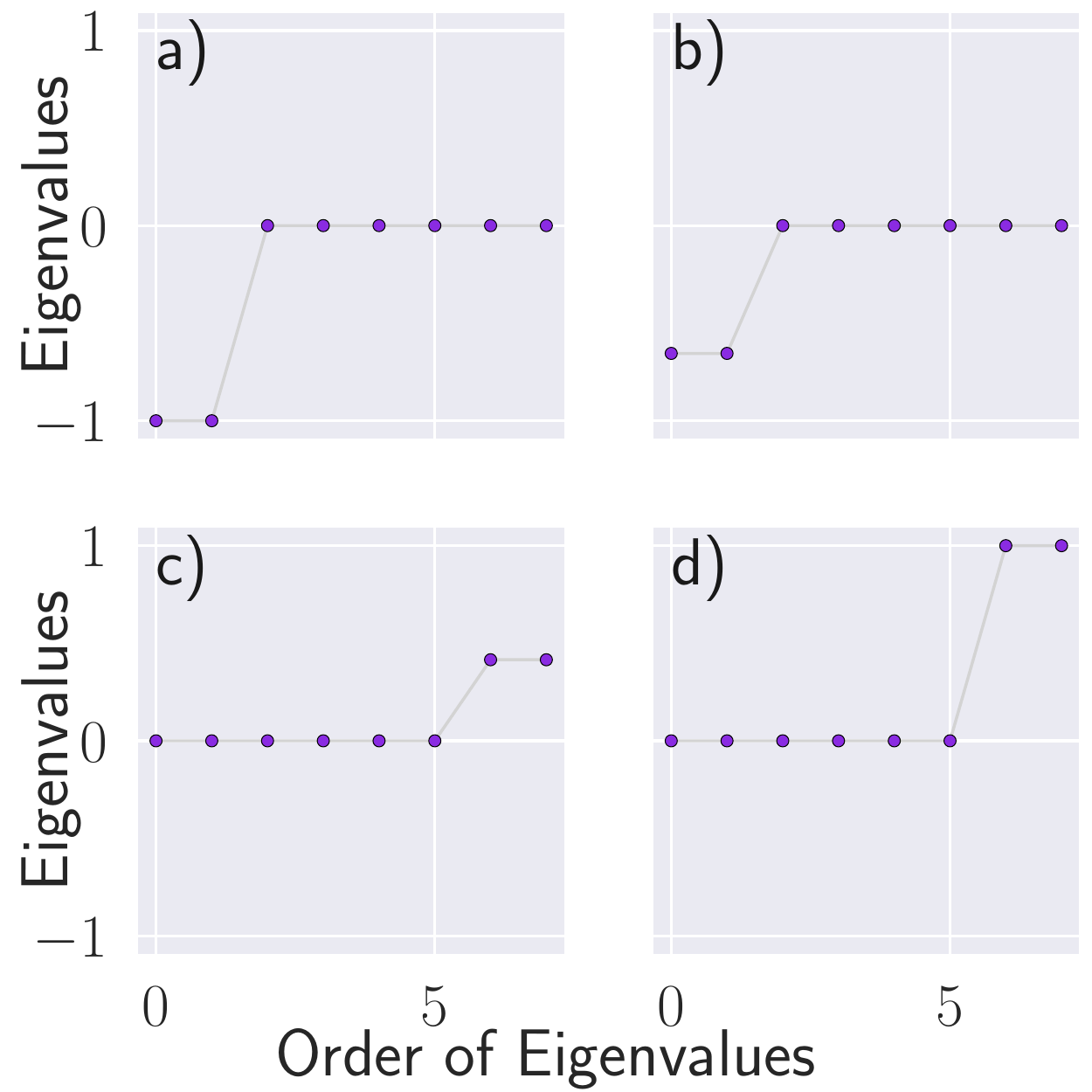}
  \end{minipage}
      \caption{\textbf{Loss landscape of the toy model with a local loss function.}  We label in the contour plot the points for which we calculate the Hessian and its eigenvalues.  The loss landscape is shown here for only 2 qubits, because for the local loss if we have more than 2 qubits and fix their rotational parameters $\bm \theta$ for the 3D loss, we will not obtain the full range variation of the loss from 0 to 1.  The landscape looks qualitatively the same for more qubits.}
 \label{fig:Loss_landscape_toy_local_loss} 
\end{figure}

Figure~\ref{fig:Loss_landscape_toy_local_loss} shows the loss landscape for a local loss function $l = 1 - \sum_i \vert \braket{\Psi |0}_i \vert^2 $, where $\ket{\Psi} = V(\boldsymbol{\theta}) \ket{\boldsymbol{0}}$ is the variational state and $\ket{0}_i$ is the qubit state of qubit $i$. In contrast to the global loss of Eq.~\eqref{Eq: Loss global Toy model}, we measure each qubit separately. For the local loss, we find that there is no point in the loss landscape with vanishing eigenvalues of the Hessian .

\begin{figure}
  \centering
  \begin{minipage}[b]{0.23\textwidth}
    \includegraphics[width=\textwidth]{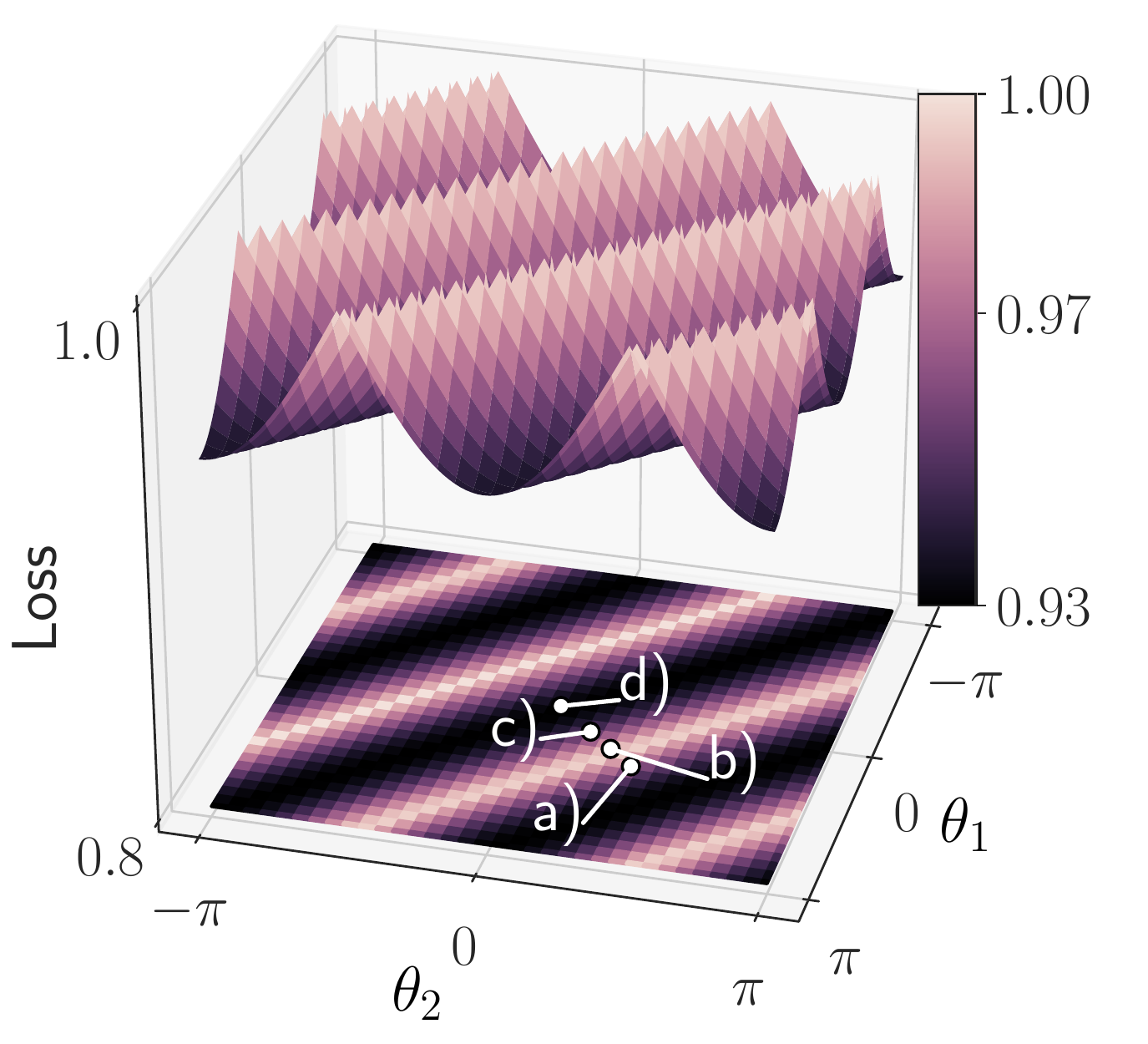}
  \end{minipage}
  \hfill
  \begin{minipage}[b]{0.23\textwidth}
    \includegraphics[width=\textwidth]{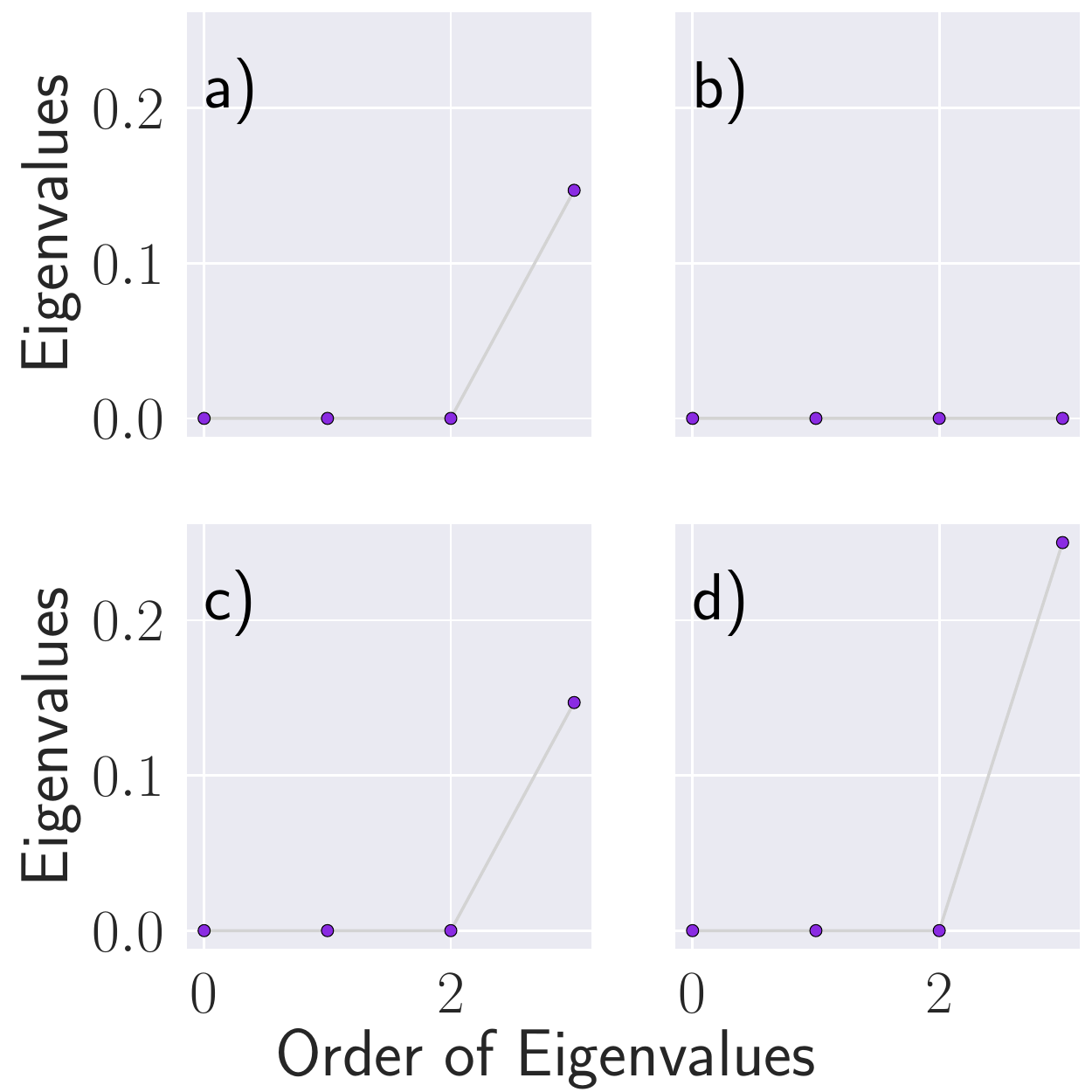}
  \end{minipage}
      \caption{\textbf{Loss landscape of the toy model with $\ket{\psi_T} = \sum_{\{\boldsymbol{\sigma} \}} \ket{\boldsymbol{\sigma}}$.} In this case, the circuit is under-parametrized and cannot reach minimum loss. Nevertheless, we can read from the Hessian's eigenvalues that we reached a stable minimum. Furthermore, one of the eigenvectors of the Hessian with eigenvalue $\lambda = 0$ is  $\bm{v}=(\theta_1,-\theta_2)$. Along this direction, the loss is constant. The latter can be verified in the contour plot.}
      \label{fig:Loss_landscape_toy_non_trivial_target} 
\end{figure}

Note that this behaviour also applies to circuit ansatzes that do not contain the exact target state and converge at high losses. If we choose the equal superposition as a target state $\ket{\psi_T} = \sum_{\{\boldsymbol{\sigma} \}} \ket{\boldsymbol{\sigma}}$ the circuit $V(\boldsymbol{\theta})$ that only consists of Pauli X rotations will not be able to rotate the initial state $\ket{0}$ such that it matches the target state with fidelity $1$. However, with an eigenvalue pattern like in Fig.~\ref{fig:Loss_landscape_toy_non_trivial_target}, we know that point d) is a local minimum, given that all the eigenvalues are either positive or zero. It becomes also clear that zero eigenvalues of the Hessian correspond to directions where changes in parameters do not affect the loss landscape. This phenomenon is illustrated in Figure~\ref{fig:Loss_landscape_toy_non_trivial_target} the direction $(\theta_1,-\theta_2)$ moves along a valley with constant loss. Therefore, the eigenvectors of the Hessian reveal additional information which help to find directions in the loss landscape of maximum or minimum stability. The latter might be used for pruning the network~\cite{NIPS1989_250}.

\section{Computation of the Hessian of a quantum circuit}
\label{sec:Hessian_of_QC}

Like in classical machine learning, we can treat the VQC as a black-box function that takes classical data $\Vec{x}$, depends on some learnable parameters $\boldsymbol{\theta}$ and returns some classical value $f(\boldsymbol{\theta}, \Vec{x})$ from a measurement or some expectation value. The gradient of any quantum circuit can be calculated by estimating the central finite difference of the circuit output 
\begin{equation}
\partial_{\theta_i} f(\boldsymbol{\theta}, \Vec{x})= \lim_{\varepsilon_i \xrightarrow{} 0}\frac{ f(\boldsymbol{\theta}_{\neg i}, \theta_i + \varepsilon_i, \Vec{x}) - f(\boldsymbol{\theta}_{\neg i}, \theta_i - \varepsilon_i, \Vec{x})}{2 \varepsilon_i},
\end{equation}
where $\boldsymbol{\theta}_{\neg i}$ are all the parameters except $\theta_i$.
\begin{figure}[t]
  \includegraphics[width=\linewidth]{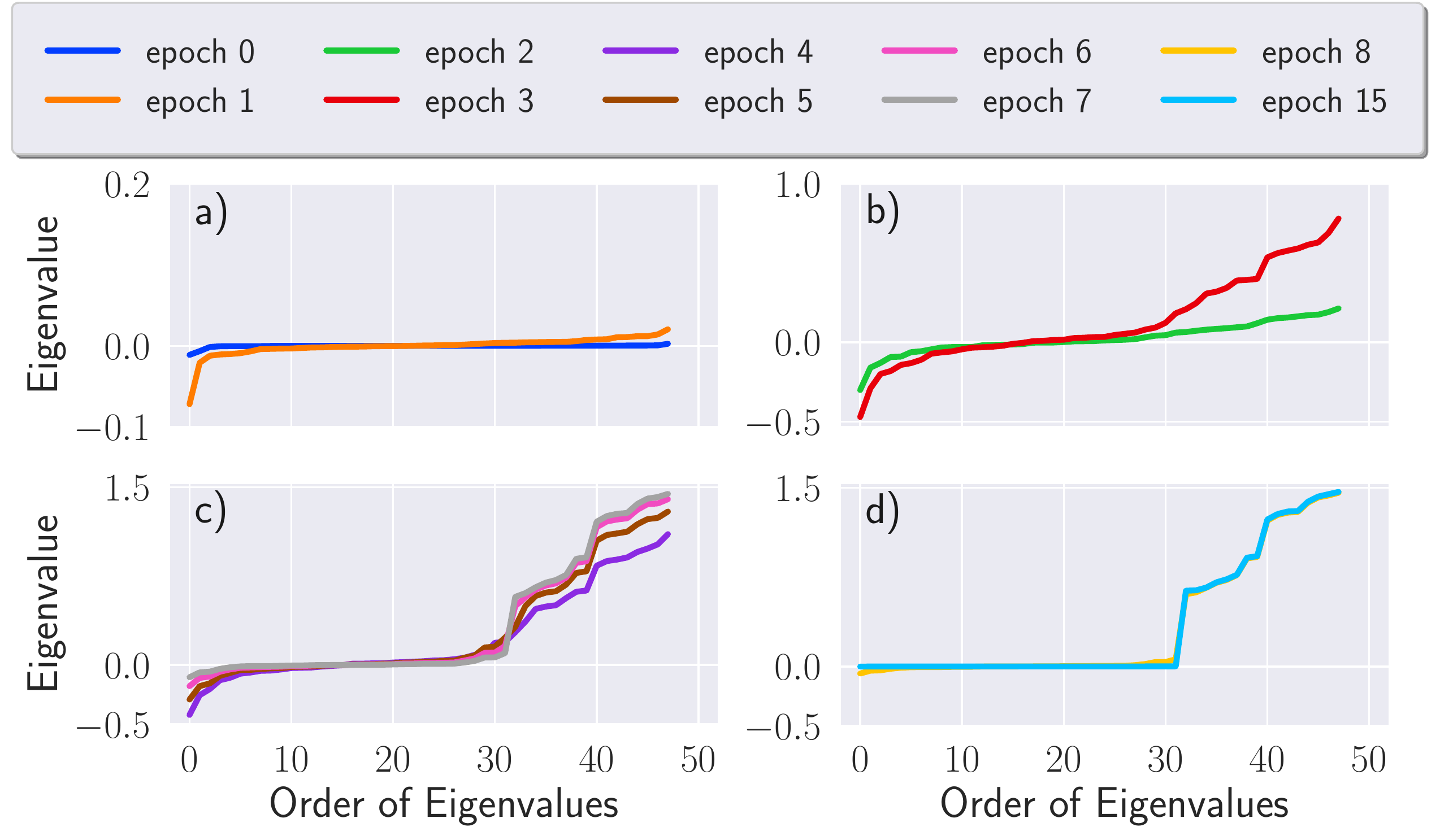}
  \caption{\textbf{Eigenvalue evolution during the training.} We separate the curves for different epochs to increase the visibility as the difference between the smallest and biggest eigenvalues is varying during the training.}
  \label{fig:complex_circ_EV_evolution}
\end{figure}
To approximate the limit of $\varepsilon_i \rightarrow 0$ on real hardware, one should choose $\varepsilon \ll 1$. This kind of gradient estimation has been shown to not perform well and depends strongly on the stochasticity of the measurement outcomes, the number of measurements and the choice of~$\varepsilon$ \cite{brekelmansGradientEstimationSchemes2005}. Because of the factor $\frac{1}{\varepsilon}$, the measurement noise can be amplified and the estimation of the gradient requires more measurement shots for smaller $\varepsilon$. In Refs.~\cite{mitaraiQuantumCircuitLearning2018,schuldEvaluatingAnalyticGradients2019}, the authors showed how the gradient can be calculated analytically to avoid the dependency on the hyperparameter $\varepsilon$. 
The authors in \cite{schuldEvaluatingAnalyticGradients2019} focused on quantum circuits that can be implemented on NISQ devices, where the parametrized gates are any qubit rotations generated by Pauli operators. Taking  advantage of these gates, the analytic gradient reads
\begin{align}
\partial_{\theta_i} f(\boldsymbol{\theta}, \Vec{x}) = \frac{1}{2} \bigg[ \langle f(\boldsymbol{\theta}_{\neg i}, \theta_i + s, \Vec{x})\rangle - \langle f(\boldsymbol{\theta}_{\neg i}, \theta_i - s, \Vec{x})\rangle \bigg],
\end{align}
where we set $s=\pi/2$. This is called the parameter shift rule and $s=\pi/2$ for all qubit gates that are generated by matrices with eigenvalues equal to $\pm 1/2$, for example Pauli spin-1/2 matrices. To obtain the gradient of a loss function $l(f(\boldsymbol{\theta}, \Vec{x}))$, we apply the chain rule $\partial_{\theta_i} l( f(\boldsymbol{\theta}, \Vec{x})) = \partial_{\theta_i} f(\boldsymbol{\theta}, \Vec{x}) l'( f(\boldsymbol{\theta}, \Vec{x})) $, where $l'$ is the first derivative of the loss function with respect to the argument $f(\boldsymbol{\theta}, \Vec{x})$.

As proposed in \cite{mitarai2019methodology}, the Hessian of a quantum circuit can be computed by applying the parameter shift rule twice
\begin{align}
\partial_{\theta_j}\partial_{\theta_i} f(\boldsymbol{\theta}, \Vec{x}) = &\frac{1}{4}~  \bigg[\langle f(\boldsymbol{\theta}_{\neg i,j}, \theta_i + s, \theta_j + s, \Vec{x})\rangle \nonumber \\
&+ \langle f(\boldsymbol{\theta}_{\neg i,j}, \theta_i - s, \theta_j - s, \Vec{x})\rangle \nonumber \\ 
&-\langle f(\boldsymbol{\theta}_{\neg i,j}, \theta_i - s, \theta_j + s, \Vec{x})\rangle \nonumber\\
&-\langle f(\boldsymbol{\theta}_{\neg i,j}, \theta_i + s, \theta_j - s, \Vec{x})\rangle \bigg].
\end{align}
With this tool we are able to study the curvature of a loss function $l(f(\boldsymbol{\theta}, \vec{x}), y)$ via the second derivative of the loss. We emphasize here that one has to apply the chain rule twice to obtain the correct Hessian of $l$
\begin{align}
\partial_{\theta_j}\partial_{\theta_i} l(f(\boldsymbol{\theta}, \vec{x}), y) &= \partial_{\theta_j}\partial_{\theta_i} f(\boldsymbol{\theta}, \Vec{x}) l'(f(\boldsymbol{\theta}, \Vec{x})) \nonumber\\
&+ \partial_{\theta_i} f(\boldsymbol{\theta}, \Vec{x}) \partial_{\theta_j}f(\boldsymbol{\theta}, \Vec{x}) l''(f(\boldsymbol{\theta}, \Vec{x})),
\end{align}
where $l'$ ($l''$) is the first (second) derivative of the loss function with respect to the argument $f(\boldsymbol{\theta}, \vec{x})$. In this work, we mostly use the loss function $l(f(\boldsymbol{\theta}, \Vec{x})) = 1 - f(\boldsymbol{\theta}, \Vec{x})$, because $f(\boldsymbol{\theta}, \Vec{x})$ will be a figure of merit to be maximized, such as the overlap with some target state or a local observable. The only exception is the training on classical data where we use the square loss to compare the labels with an observable.

\section{Behaviour of a general VQC without data}
\label{Sec:General_Circuit}
\begin{figure}[t]
  \includegraphics[width=\linewidth]{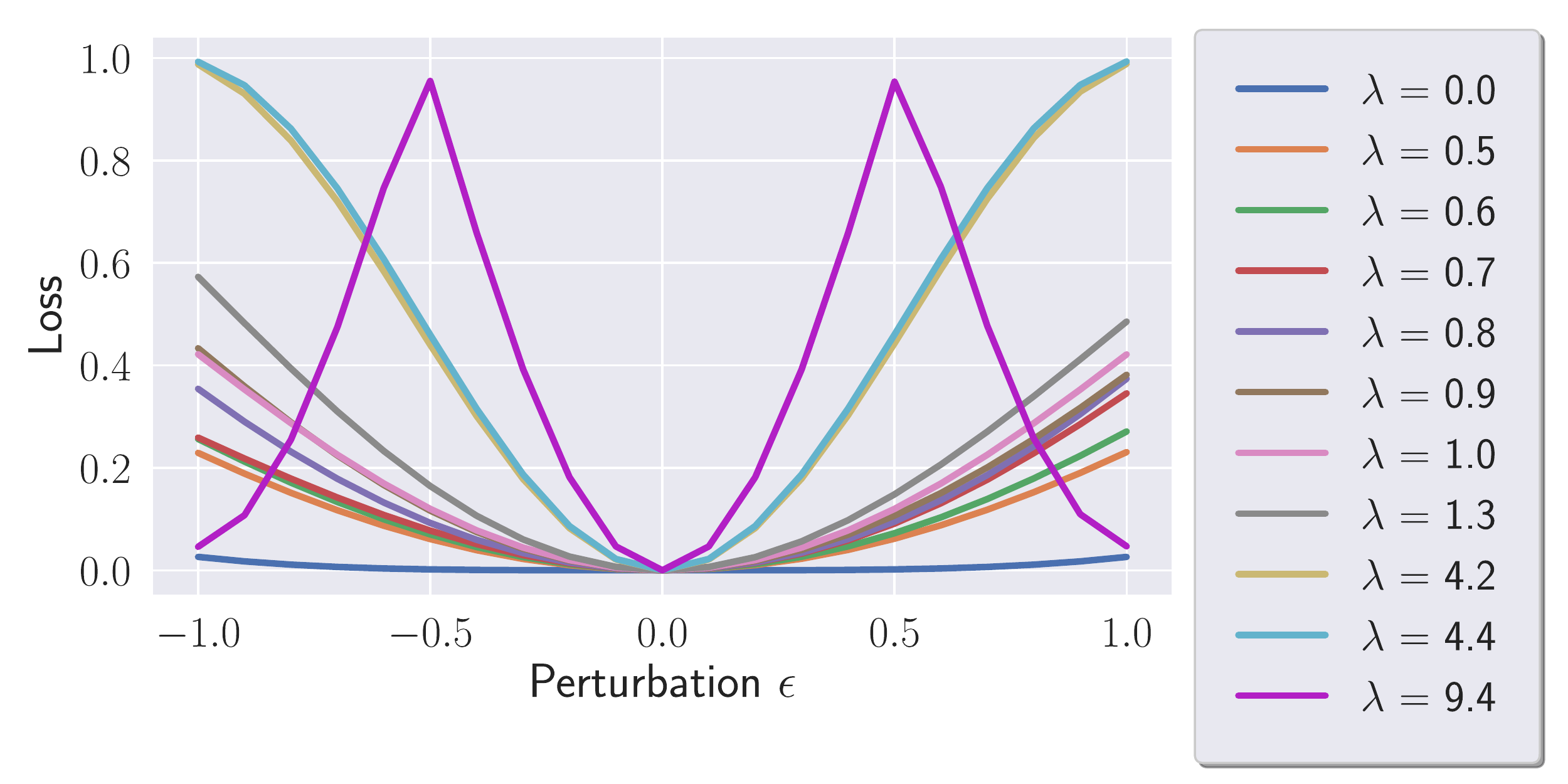}
  \caption{\textbf{Loss around the mimimum for a given pertubation $\epsilon$.} We show several perturbations $\boldsymbol{\theta}_{\text{pert}} = \boldsymbol{\tilde{\theta}} + \epsilon \, \bf{v}$ labeled by the eigenvalue $\lambda$ and the perturbation is along $\bf{v}$ which is the corresponding eigenvector of the eigenpair $(\lambda,\bf{v})$. Perturbations in the direction of a zero eigenvalue behave similarly and therefore we only show one as an example. Perturbations along the eigenvector of big eigenvalues lead to a much steeper increase in the loss.}
  \label{fig:perturbation_loss_minima}
\end{figure}

We now focus on the behaviour of the eigenvalues of the Hessian during the training of a more general circuit with entangling gates. The aim of the training is, again, to optimize a randomly initialized circuit $V(\bm \theta)$ such that we obtain the target state $\ket{\psi_T} = \sum_{\{\boldsymbol{\sigma} \}} \ket{\boldsymbol{\sigma}}$ at the output. To have a better circuit ansatz than in the previous section, we use general qubit rotations $R(\phi_1, \phi_2, \phi_3) = e^{i \phi_1 \sigma_z}e^{i \phi_2 \sigma_y}e^{i \phi_3 \sigma_z}$, which we apply to each qubit, followed by a layer of controlled-Z (CZ) operations. We entangle the gates following the architecture proposed in Ref.~\cite{perez-salinasDataReuploadingUniversal2020}, where after each layer of rotations we entangle each qubit with its neighbouring qubit, starting either with the even or the odd qubits. We define the layer $L_i(\boldsymbol{\theta}_i)$ as the sequence of rotating operations followed by CZ operations. Depending on the even/odd label $i$ of the layer, we start the pairwise entanglement with the even/odd qubits [See Fig.~\ref{fig:Toy_and_General_Circuit} (b)]. 

Figure~\ref{fig:complex_circ_EV_evolution} shows the typical behaviour of the eigenvalues of the Hessian of such a parametrized circuit during the training. The eigenvalues' distribution for the randomly initialized circuit (at epoch 0) shows the occurrence of a Barren-Plateau-like behaviour with all eigenvalues close to zero. Already for a circuit with 4 qubits and 4 layers, which has 48 parameters (3 for each rotation), the circuit is likely to be initialized in a flat region.

In particular, Figure~\ref{fig:complex_circ_EV_evolution}~(d) depicts the eigenvalues of the Hessian for the completely converged variational circuit (Epoch 15) and, similarly to the previous toy model, the eigenvalues are all non negative. Furthermore, most of the eigenvalues are zero, which means that the minimum is a flat pool, with only a few directions where the loss will increase and where most directions in the loss landscape are unaffected by small perturbations.

3D visualizations in these high dimensional loss landscapes are not feasible anymore, provided that it is not clear how to fix the remaining parameters. However, the Hessian contains information about the local curvature of the loss landscape. Considering an eigenvector $\bf{v}_0$ of the Hessian after convergence with the corresponding eigenvalue $\lambda_0=0$, we know that if we perturb these optimal parameters $\boldsymbol{\tilde{\theta}}$ in the direction of $\bf{v}_0$, $\boldsymbol{\theta}_{\text{pert}} = \boldsymbol{\tilde{\theta}} + \epsilon \, \bf{v}_0$, the loss will not change for a normalized $\bf{v}_0$ and a small perturbation $\epsilon \in \mathbb{R}$.
The same idea applies to eigenvectors $\bf{v}_{\lambda}$ that correspond to an eigenvalue $\lambda \gg 0$. Perturbing $\boldsymbol{\tilde{\theta}}$ in the direction of $\bf{v}_{\lambda}$ the loss increases fast, because the loss landscape in this direction is steep.

Figure~\ref{fig:perturbation_loss_minima} shows the perturbation in the direction of several eigenvectors. The corresponding eigenvalues are the labels of the curves. For the eigenvalue $\lambda = 0$ we see that the loss only increases for perturbations $\epsilon \gg 0$, which shows that the loss landscape is indeed flat around the minimum. We also would like to emphasize that most eigenvalues $\lambda=0$, but we are only able to show the perturbation curve for one eigenpair $(\lambda_i=0, \bf{v}_i)$ in Fig.~\ref{fig:perturbation_loss_minima} because they behave very similarly around $\epsilon \ll 0$. The vast number of zero eigenvalues indicate that the loss landscape around the minimum is mostly flat with only a few steep directions.

\begin{figure}[t]
  \includegraphics[width=0.7\linewidth]{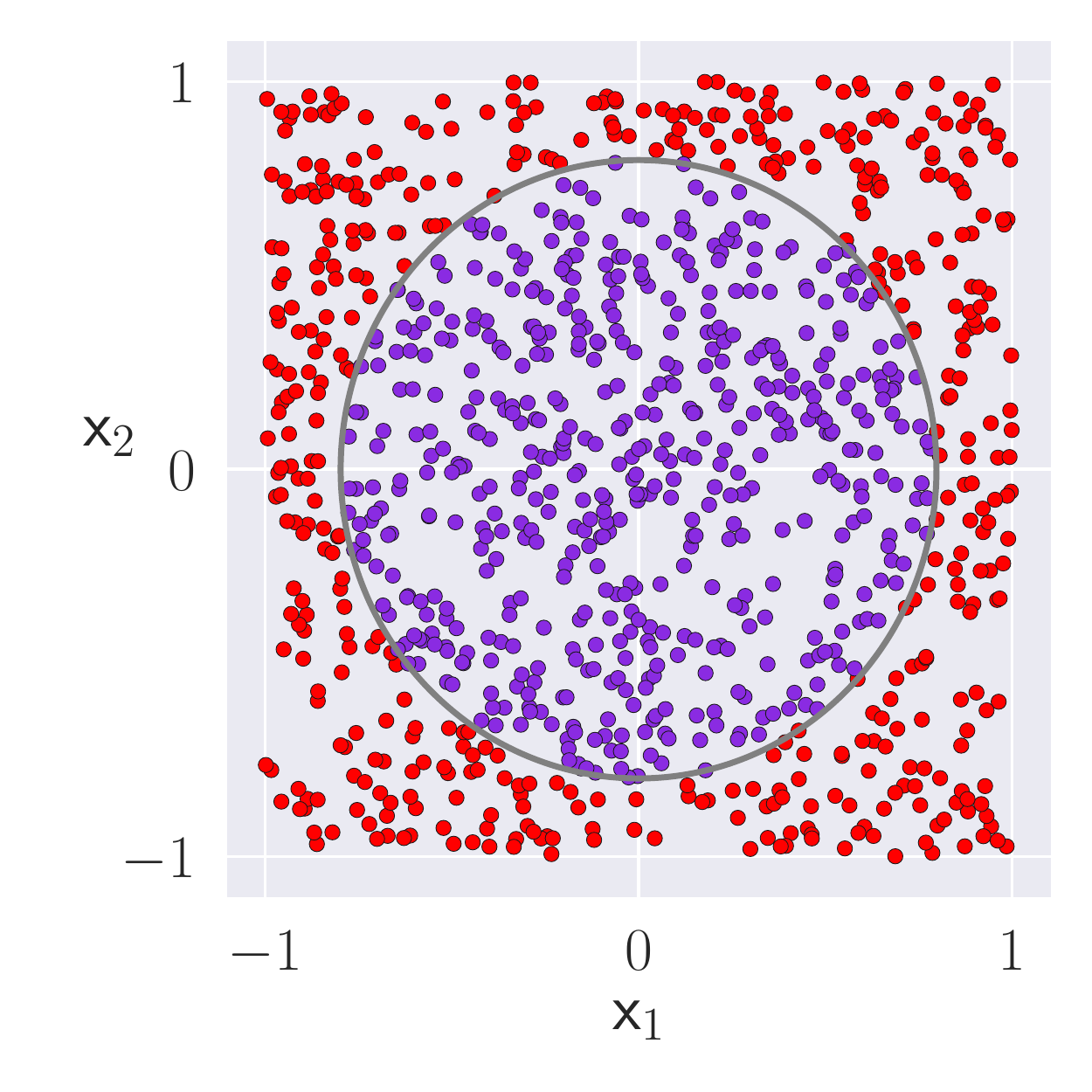}
  \caption{\textbf{Training data for variational circuit.} Data inside the disk are labelled $-1$ and data outside the disk are labelled $+1$.}
  \label{fig:training_data}
\end{figure}

We therefore see that the positive semi-definite nature of the Hessian is an indicator for a very stable solution. The optimization landscape cannot be perturbed and it is unlikely that during the optimization one jumps to another local minimum. The fact that in all directions the loss either stays the same or increases means that, with gradient descent methods the minimum loss can always be recovered, because we are in a locally convex point of the landscape. 
On the other hand, as we discuss in the next section, in quantum and classical ML tasks involving training with data not all eigenvalues are positive. There are still some small negative eigenvalues remaining and the Hessian is not positive semi-definite~\cite{sagunEigenvaluesHessianDeep2017}. The optimization still converges, but the exact nature of the loss minima is controversial. There are claims that these flat regions are essential for the generalization capabilities of a neural network~\cite{keskarLargeBatchTrainingDeep2017a} and that through these wide flat pools two solutions can be connected, which contradicts the idea of isolated basins at the bottom of the loss landscape~\cite{chaudhariEntropySGDBiasingGradient2017}.

\section{Training with data}
\label{sec:QNN_data}
We now characterize the Hessian of VQCs trained on data $X$ with labels $y$, often called quantum neural networks (QNNs). We compare the results to a classical fully connected feed forward NN and we observe qualitative differences between classical NNs and QNNs. To this end, we use a simple dataset $\mathcal{D} =  \{(\vec{x}^{\alpha}, y^{\alpha}) \}_{\alpha=1}^N$ which only contains two classes, depicted in Fig.~\ref{fig:training_data}: a 2 dimensional set of points $\vec{x}=(x_1, x_2)$ labeled $-1$ inside the circle of radius $r$ and $1$ otherwise. In order to have a balanced dataset between the labels for $x_i \in [-1,1]$, we choose $r=\sqrt{2/\pi}$.

\begin{figure}[t]
  \includegraphics[width=\linewidth]{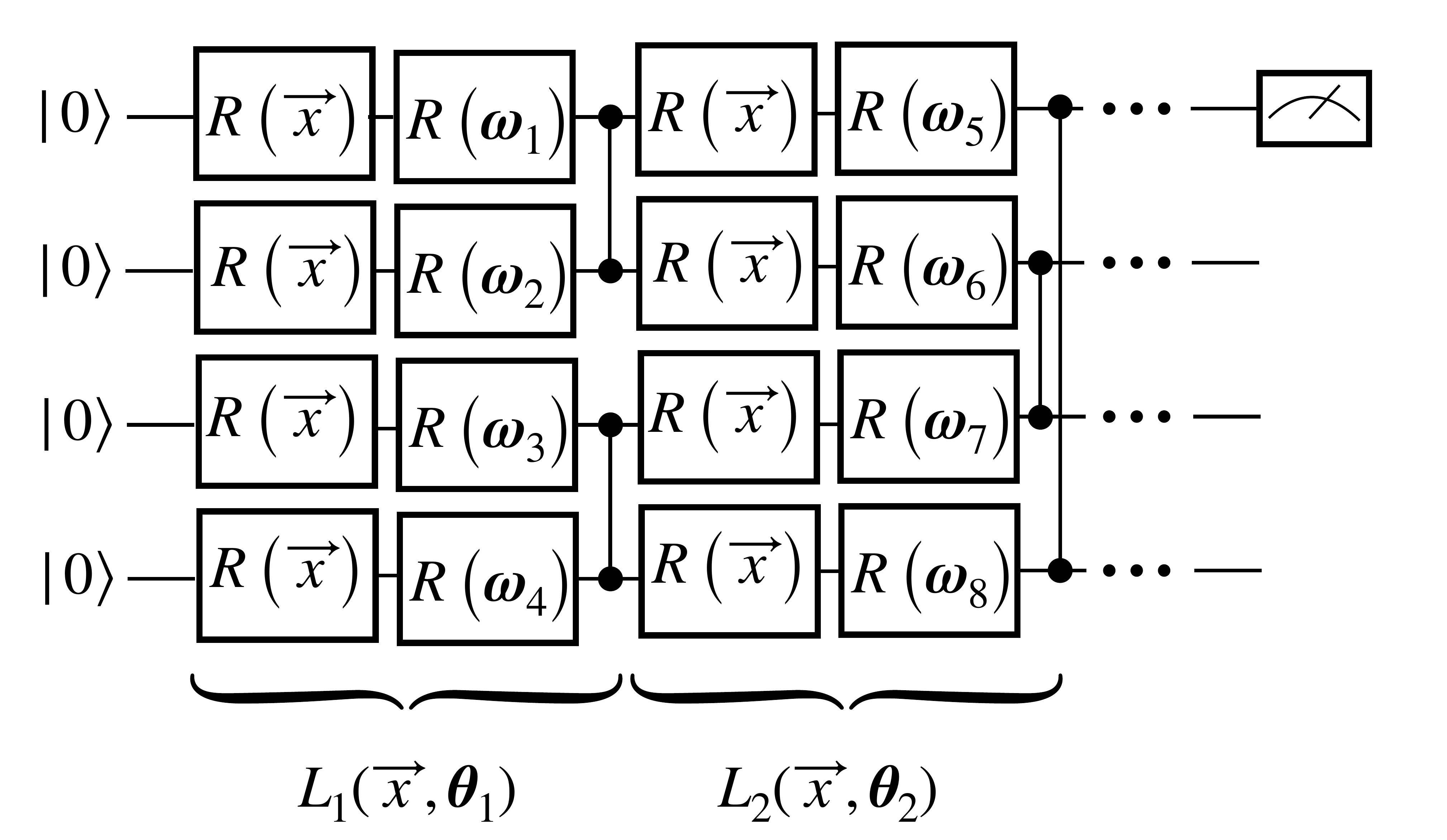}
  \caption{\textbf{Data Reuploading Circuit.} Each layer $L_i$ contains an additional qubit rotation $R(\vec{x})$ with the classical data vector $\vec{x} = (x_1, x_2, x_3)$. The parameterization and the entangling gates are the same as in Figure \ref{fig:Toy_and_General_Circuit} (b). For the 2D input data $(x_1, x_2)$, we set $x_3 = 0$. }
  \label{fig:Reuploading_Circuit}
\end{figure}

The first step is to encode the data instances $\vec{x} \in X$ onto the variational circuit and then the second step is to associate a certain measurement direction with a classical label $y$. There are many possible ways to encode the classical data in the quantum state $\ket{\psi(\bm \theta, \vec{x})}$. We follow here the architecture proposed in Ref.~\cite{perez-salinasDataReuploadingUniversal2020} and depicted in Fig.~\ref{fig:Reuploading_Circuit}, which adds to each layer of Sec.~\ref{Sec:General_Circuit} an additional rotation $R(x_1, x_2, x_3)$. This way, each layer consists of a data dependent  rotation $R(x_1, x_2, x_3)$ followed by a parametrized rotation $R(\omega_1, \omega_2, \omega_3)$ applied on each qubit (we here choose $x_3=0$ as the data are two-dimensional), followed by the entangling operators. Regarding the label, the general idea is to define a measurement on one or more of the qubits that determines the QNN prediction. For two classes, the measurement of one qubit is already sufficient to have orthogonal measurements for each class, minimizing the overlap between complementary predictions. Therefore we choose to measure the first qubit in the Pauli $Z$ direction and associate the measurement expectation value $\langle Z \rangle = 1$ ($\langle Z \rangle = -1$) with the labels $y=1$ (y=-1). The loss $l(\braket{Z}, y)$ compares the label with the prediction $\braket{Z}$ and we here use the mean square loss 
\begin{align}
l(\vec{x}, y) = (\bra{\psi(\bm \theta, \vec{x})} Z \ket{\psi(\bm \theta, \vec{x})} - y)^2.
\label{eq:square_loss_Z}
\end{align}
%
After training, one finds the set of parameters $\bm{\tilde{\theta}}$ that minimizes the emprical risk $\mathcal{L} = \sum_{\alpha} l(\vec{x}^{\alpha},y^{\alpha})$. 

To get an intuition, we first consider the Hessian of the loss of an arbitrary single training point $(\vec{x},y)$. It has an eigenvalue distribution dependent on the prediction of the NN on $\vec{x}$. A correctly predicted label $\hat{y}$ with a loss $l(\vec{x}, \hat{y} = y)$ will lead to an eigenvalue distribution with most eigenvalues equal 0 and a few bigger than zero. This is the typical distribution if the loss of the NN is in a minimum \cite{sagunEmpiricalAnalysisHessian2018}. For wrongly predicted labels, the eigenvalues of $H$ are mostly zero with a few negative values. This is typical if the loss of the QNN is in a maximum. If the prediction $\hat{y}$ is somewhere between the labels $-1$ and $1$ and there is a big uncertainty in the QNN's prediction, the eigenvalues of the Hessian will be again mostly zero but with an equal amount of positive and negative eigenvalues. This distribution is likely to be obtained in randomly initialized NNs. In the following, we show that for VQCs the eigenvalue distribution of the Hessian and the loss landscape itself shows some qualitative differences with respect to classical neural networks. We also emphasize that we do not look at the Hessian spectrum of single training points, but rather at the spectrum of the whole training set. Empirical studies show that the eigenvalues of the sum of all the losses of the single training points (the empirical risk) behave approximately like the average of the spectrum of the single losses. If most training points are correctly classified, the Hessian of the empirical risk has negligible negative eigenvalues. In contrast, if all points are incorrectly classified the positive eigenvalues are negligible, if most training points show uncertainty in their prediction then there will be a a positive and a negative tail in the eigenvalue spectrum.

\begin{figure}[H]
  \centering
  \begin{minipage}[b]{0.22\textwidth}
    \includegraphics[width=\textwidth]{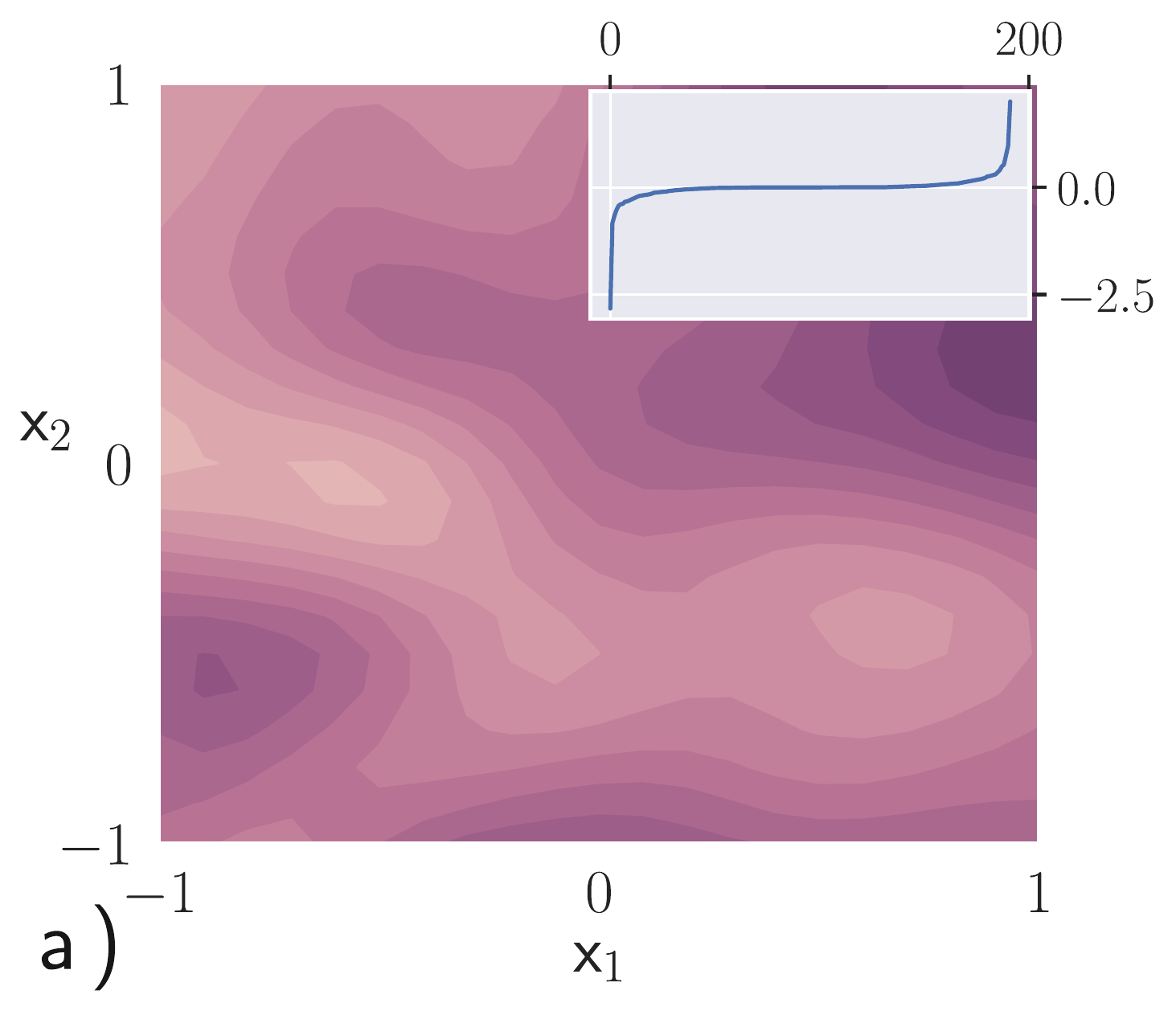}
  \end{minipage}
  \hfill
  \begin{minipage}[b]{0.255\textwidth}
    \includegraphics[width=\textwidth]{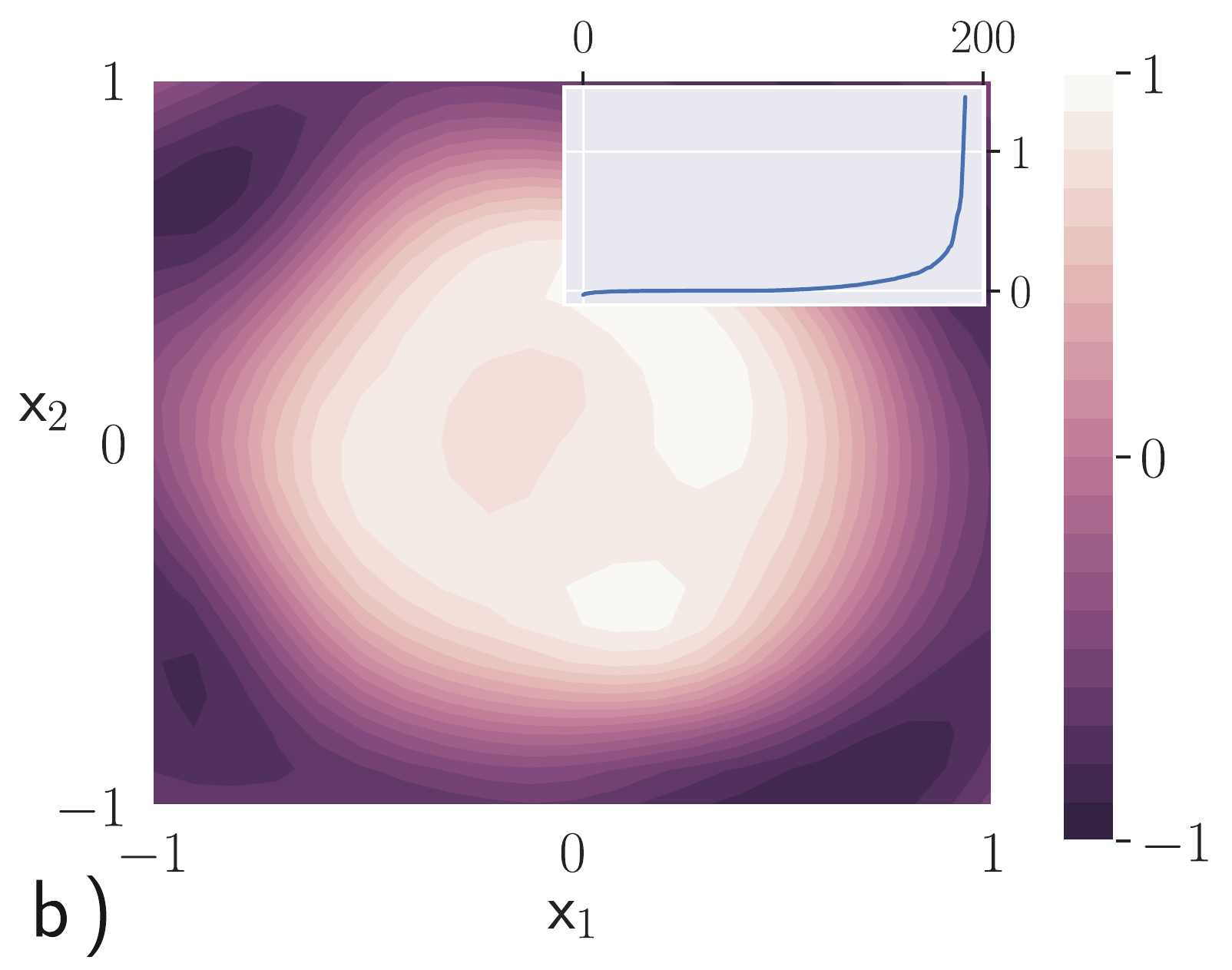}
  \end{minipage}
  \caption{\textbf{QNN label prediction map  with \emph{Z}-measurement} for a randomly initialized variational circuit a) and after the training b). The inset shows the Hessian's eigenvalue distribution, which is similar to a classical NN if we use the loss function . The Hessian is computed over the whole training set.}
    \label{fig:QNN_pred_Z}
\end{figure}

\begin{figure}[t]
  \centering
  \begin{minipage}[b]{0.22\textwidth}
    \includegraphics[width=\textwidth]{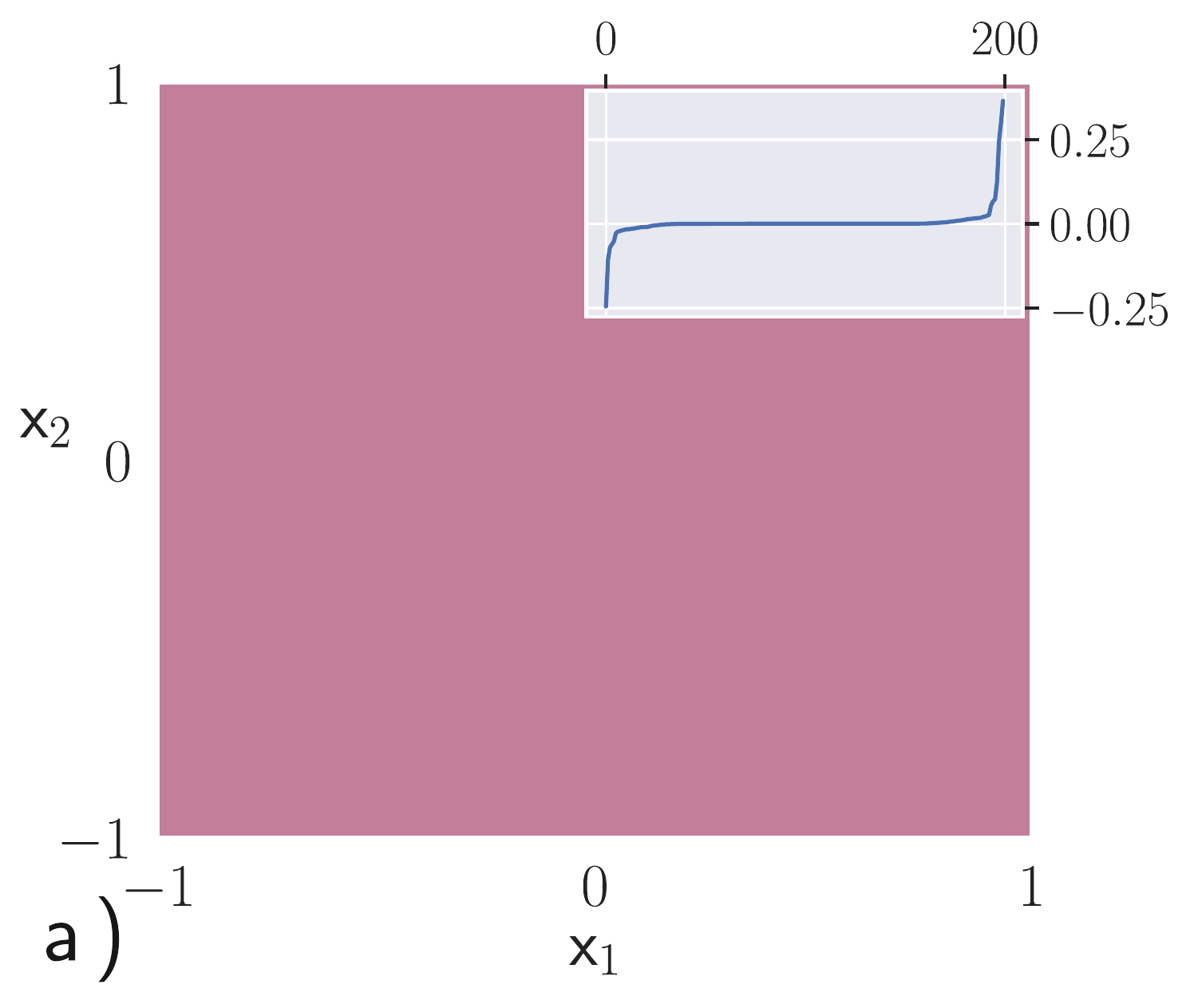}
  \end{minipage}
  \hfill
  \begin{minipage}[b]{0.25\textwidth}
    \includegraphics[width=\textwidth]{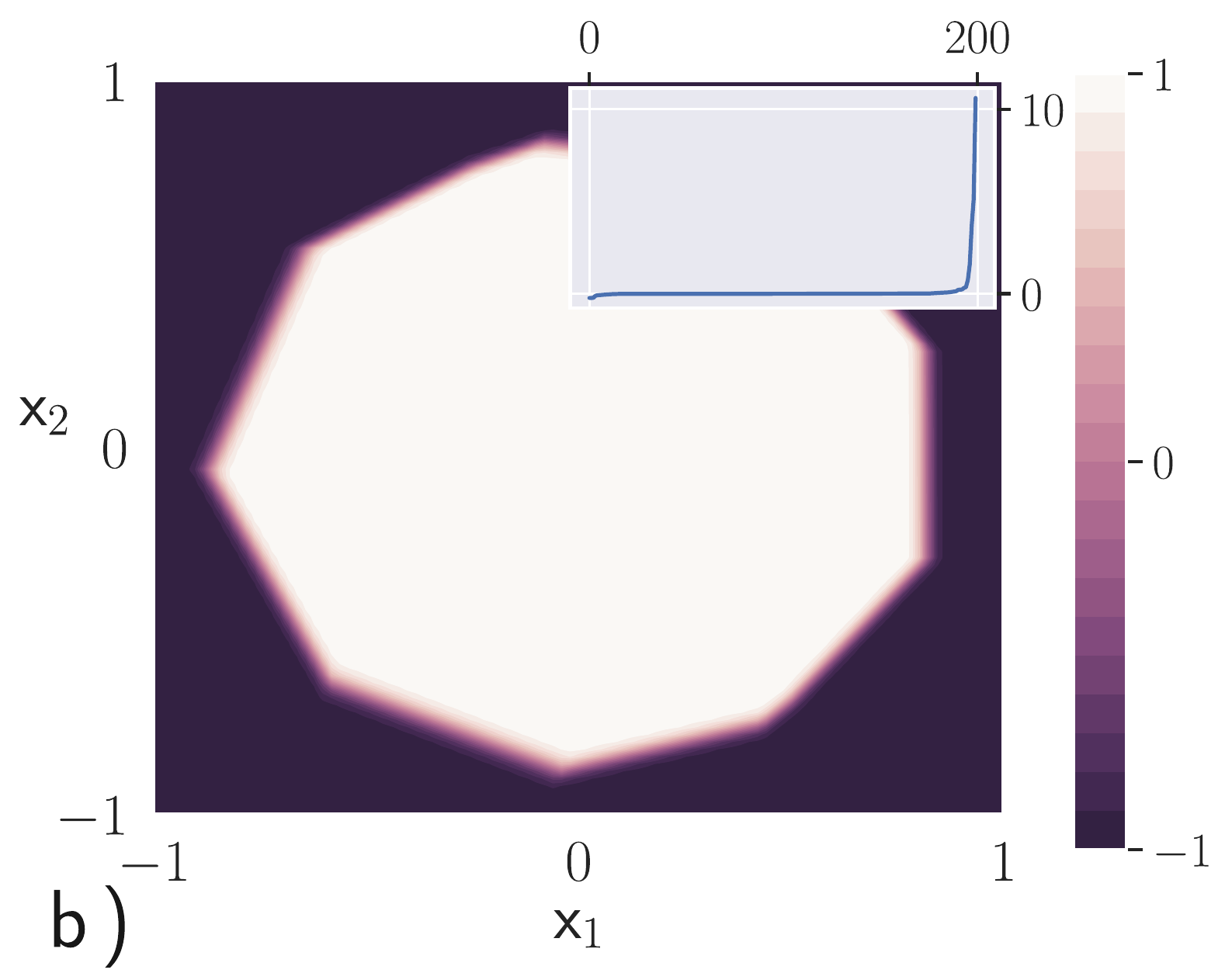}
  \end{minipage}
  \caption{\textbf{FFNN label prediction map} for a randomly initialized NN  a) and after the training b). The inset shows the Hessian's eigenvalue distribution. The Hessian is calculated over the whole training set.}
    \label{fig:CLNN_pred}
\end{figure}

Figure \ref{fig:QNN_pred_Z} shows the prediction map of the QNN and the eigenvalues of the Hessian calculated over the whole training set (inset) for random initialization and after the training. We compare the results of the QNN with a classical fully connected feed forward NN (FFNN) in Figure \ref{fig:CLNN_pred}. To make the results comparable, we choose a NN with a comparable amount of parameters, the same labels $y= \pm 1$ and the same loss function. The classical NN has 2 hidden layers with 12 and 10 neurons which results in 200 parameters, which is comparable to the 192 parameters of the QNN (8 qubits, with each 8 layers and 3 parameters per layer $8 \times 8 \times 3 = 192$).

The distribution of the eigenvalues is similar for the FFNN and the QNN. For a random initialization, the eigenvalues are equally likely positive and negative and the bulk is zero. After convergence, the negative eigenvalues disappear. There is a qualitative difference between the classical and the quantum version concerning the amplitude of the eigenvalues, which is almost an order of magnitude smaller for the FFNN for random initialization. This effect persists for different architectures, with different numbers of hidden layers or nodes and as well for higher overparametrization, but it depends on the activation function of the NN.  We here use the mean square loss.
For these relatively small FFNNs with 200 parameters, the distribution of the eigenvalues for a random initialization can vary strongly, but this is an effect that disappears for higher overparameterization of the FFNN. This is in contrast to the QNN that does not show strong fluctuations in the amplitude of the negative and positive eigenvalues. 

Furthermore, the loss landscape of the QNN already shows steep slopes for random initialization compared to the flat random initialization of the FFNN. This seems to come from the fact that the loss of a QNN contains products and sums of sinusoidal  functions like in Eq.~\eqref{Eq:General_Target}, and stands in stark contrast to the training of variational circuits without data, where it is likely to initialize the circuit in absolutely flat regions, the Barren Plateaus. We did not observe any Plateaus in the case of QNN trained on data.

As the authors in Ref.~\cite{wierichsAvoidingLocalMinima2020b} pointed out, classical NNs tend to transform local minima into deep and narrow steep valleys \cite{liBenefitWidthNeural2020}, which means that, for FFNN the amplitude of the negative eigenvalues tends to decrease if the training progresses. This is also in agreement with Ref.~\cite{alainNegativeEigenvaluesHessian2019}. 
In addition, for  FFNNs, we  observe that there are just a few positive eigenvalues, of the order of the number of classes as introduced in section \ref{sec:LL_of_cl_NN}, in agreement with \cite{sagunEmpiricalAnalysisHessian2018}.
On the other side, QNNs do not seem to converge to these extremes, where just a few eigenvalues are positive and their positive "tail" in the eigenvalue spectrum is less steep (at least for the order of $<200$ parameters). This observation seems to be connected with the fact that the loss landscape of the QNN after convergence still has some uncertainty at the decision boundary, unlike the FFNN which barely has any. This region of uncertainty is a shortcoming of the QNN and it is because of a lack of expressive power. In \cite{schuld2020effect} the authors describe the reuploading scheme in terms of a Fourier series. Therefore, following this logic, we could approximate a stept function at the decision boundary by adding more layers of reuploading. The eigenvalues of the Hessian of the loss landscape inside a pool with high certainty are all close to zero, even if the NN prediction is wrong. This is generally true for wide flat regions of the loss landscape and it means that the main contribution of the eigenvalues of the Hessian comes from transition regions in the loss landscape where the label prediction changes from one to another.

\section{Faster training convergence via the Hessian}
\label{sec:Barren_Plateau}

In this section, we show how the Hessian can be used to adjust the learning rate for faster convergence during the training of variational circuits. 

\begin{figure}[t]
  \includegraphics[width=\linewidth]{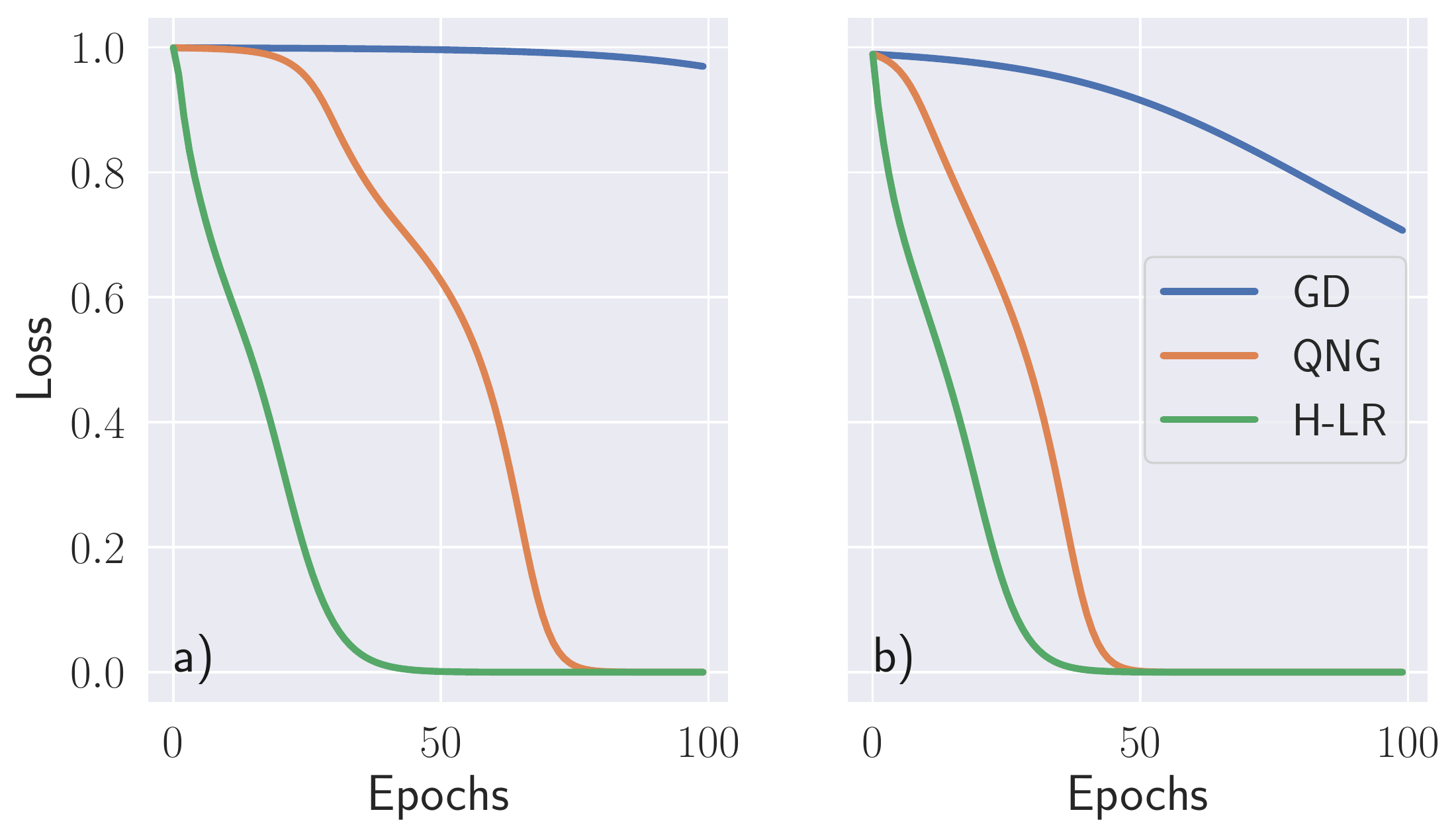}
  \caption{\textbf{Training loss for training with} Gradient descent (GD), Quantum Natural Gradient (QNG) and the learning rate adaption via the Hessian's maximum eigenvalue (H-LR) for two different random initializations a) and b). The first initialization a) shows a very slow convergence for GD and QNG in the beginning. This is caused by a Barren Plateau.} 
  \label{fig:Compare_methods}
\end{figure}

We propose to use the inverse of the largest eigenvalue of the Hessian $1/\lambda_{\text{max}}$ as the learning rate of the gradient descent optimizer in order to avoid local traps in the loss landscape. Figure~\ref{fig:Compare_methods} shows a comparsion between a normal gradient descent optimizer (GD), the quantum natural gradient optimizer (QNG) from Ref.~\cite{stokesQuantumNaturalGradient2019a} and our method of a learning rate of $1/\lambda_{\text{max}}$. The comparison has been done for random initializations of a circuit with $N=8$ qubits and 4 layers of parameterized rotations $R(\phi, \theta, \omega)$. The initial state is $\ket{0}^{\otimes N}$ and the target state is $\ket{\psi_T} = 1/\sqrt{2^N} \sum_{\boldsymbol{\sigma}} \ket{\boldsymbol{\sigma}}$. To compare the optimization methods we start in the same random initialization. We observe that the GD method is already struggling to train for such small circuits, because the gradients are too small, especially at the beginning of the training. With large learning rates, we might be able to escape the region of small gradients, but it is impossible to converge. Both methods QNG and our Hessian method escape the flat region. The QNG method basically transforms the parameter update such that the update direction is not only in the steepest direction of the Euclidean space of the parameters, but in the steepest direction of the distribution space of all possible loss functions. Nevertheless, QNG can still get stuck in a flat region of the loss function. Hence, QNG provides poor performance whenever the circuit is initialized in a region with small gradients. Our analysis also explains, why LBFGS optimization methods work better in ref.~\cite{perez-salinasDataReuploadingUniversal2020}, provided that they are Hessian based. 

Furthermore Ref.~\cite{wierichsAvoidingLocalMinima2020b} shows that natural gradient optimizers help to prevent falling into local minima whereas Hessian based methods struggle with local minima. The full Hessian is also costlier to calculate than the quantum metric tensor from Ref.~\cite{stokesQuantumNaturalGradient2019a}. Hence, we propose to use our method to escape from a flat region of the loss landscape, but to use the QNG for regions where the gradients are bigger. The two methods can, therefore, be combined: The QNG adapts the local shape of the gradient and fits it exactly to the loss landscape, providing faster convergence than vanilla GD ~\cite{stokesQuantumNaturalGradient2019a} and the Hessian helps to kick the loss out of very flat regions. 
Finally, we emphasize that this method might have a practical shortcoming:
The authors in \cite{cerezoImpactBarrenPlateaus2020} show that the the measurement noise can be a limitation for a proper evaluation of the Hessian in a Barren Plateau as the eigenvalues and the gradients are small.

\section{Methods}

For most of the implementations we use the pennylane package from Ref.~\cite{bergholmPennyLaneAutomaticDifferentiation2020}. We also wrote our own code in pytorch to speed up experiments with a big number of parameters. For the implementation we used the complex pytorch library from Ref~\cite{beachQuCumberWavefunctionReconstruction2019}. All code can be found on GitHub \cite{Huembeli_Github_2020}.

\section{Conclusion}

In this work, we introduced the Hessian as a tool to study VQCs and discussed how the interpretation the eigenvalues and eigenvectors of the Hessian can lead to a better understanding of the loss landscape of a VQC. The combination of NISQ devices and gradient descent methods is still a young field and needs a thorough study for a better understanding of its potential. Especially for data driven tasks, the use of QML is at least controversial and one needs to find in which problems the application of QML has a potential advantage over classical NNs. We identified some qualitative differences between QNN and FFNN loss landscapes, which becomes evident after their initialization where the QNN loss shows a rougher surface than the FFNN which initializes very flat. We hope we can add another piece to the already controversial discussion of what kind of basins an ideal minima should live in for NNs by adding another interesting candidate: QNNs.
The main bottleneck for the use of the Hessian in NN optimizations is its computational cost. For future research there is a need of finding good approximation schemes for quantum circuits. A possible direction could be inspired by the classical methods to approximate the Hessian vector product in $\mathcal{O}(n)$ iterations, for $n$ parameters. We could as well explore diagonal approximations of the Hessian, similar to the diagonal approximation in QNG \cite{stokesQuantumNaturalGradient2019a}. Furthermore, there are recent developments in classical ML showing that the learning rate early in SGD determines the quality of the minima found after the training~\cite{jastrzebskiBreakEvenPointOptimization2020} and big initial learning rates, like in our method with the inverse of the Hessian's biggest eigenvalue goes in this direction. Furthermore, there are Hessian based interpretability methods like the influence function~\cite{kohUnderstandingBlackboxPredictions2017b,dawidPhaseDetectionNeural2020} which one could also apply to QNNs. 
Finally, we would like to point out that there are striking similarities between the minimization of VQCs and Quantum Monte Carlo. In particular, in both scenarios, it seems advantageous to take into account the local curvature through the (quantum) metric tensor or the Hessian~\cite{Park_2020}. Thus, it would be interesting to explore the connections between these two fields.

\emph{Note.} Recently, we became aware of two other works focused on the computation of higher order derivatives in VQCs~\cite{cerezoImpactBarrenPlateaus2020, mariEstimatingGradientHigherorder2020}.

\section{Acknowledgements}

We would like to thank Josh Izaac from Xanadu who always was kind enough to help us with question concerning the Pennylane library. A big thanks also goes to Borja Requena Pozo for his  careful reading of the manuscript and constructive comments.
We acknowledge the Spanish MINECO (National Plan 15 Grant: FISICATEAMO No. FIS2016-79508-P, SEVERO OCHOA No. SEV-2015-0522, FPI), Fundacio Cellex and Mir-Puig, Generalitat de Catalunya (AGAUR Grant No. 2017 SGR 1341 and 1381, CERCA/Program), ERC AdG NOQIA, EU FEDER, European Union Regional Development Fund - ERDF Operational Program of Catalonia 2014-2020 (Operation Code: IU16-011424), MINECO-EU QUANTERA MAQS (funded by The State Research Agency (AEI) PCI2019-111828-2 / 10.13039/501100011033), and the National Science Centre, Poland-Symfonia Grant No. 2016/20/W/ST4/00314. A.D. akcnowledges the Juan de la Cierva program (IJCI-2017-33180) and and the financial support from a fellowship granted by la Caixa Foundation (fellowship code LCF/BQ/PR20/11770012). This project has received funding from the European Unions Horizon 2020 research and innovation programme under the Marie Skodowska-Curie grant agreement No 665884 (P.H.)

\bibliographystyle{apsrev4-1}
\bibliography{QNN_Proj_Alex.bib}

\begin{thebibliography}{47}%
\makeatletter
\providecommand \@ifxundefined [1]{%
 \@ifx{#1\undefined}
}%
\providecommand \@ifnum [1]{%
 \ifnum #1\expandafter \@firstoftwo
 \else \expandafter \@secondoftwo
 \fi
}%
\providecommand \@ifx [1]{%
 \ifx #1\expandafter \@firstoftwo
 \else \expandafter \@secondoftwo
 \fi
}%
\providecommand \natexlab [1]{#1}%
\providecommand \enquote  [1]{``#1''}%
\providecommand \bibnamefont  [1]{#1}%
\providecommand \bibfnamefont [1]{#1}%
\providecommand \citenamefont [1]{#1}%
\providecommand \href@noop [0]{\@secondoftwo}%
\providecommand \href [0]{\begingroup \@sanitize@url \@href}%
\providecommand \@href[1]{\@@startlink{#1}\@@href}%
\providecommand \@@href[1]{\endgroup#1\@@endlink}%
\providecommand \@sanitize@url [0]{\catcode `\\12\catcode `\$12\catcode
  `\&12\catcode `\#12\catcode `\^12\catcode `\_12\catcode `\%12\relax}%
\providecommand \@@startlink[1]{}%
\providecommand \@@endlink[0]{}%
\providecommand \url  [0]{\begingroup\@sanitize@url \@url }%
\providecommand \@url [1]{\endgroup\@href {#1}{\urlprefix }}%
\providecommand \urlprefix  [0]{URL }%
\providecommand \Eprint [0]{\href }%
\providecommand \doibase [0]{http://dx.doi.org/}%
\providecommand \selectlanguage [0]{\@gobble}%
\providecommand \bibinfo  [0]{\@secondoftwo}%
\providecommand \bibfield  [0]{\@secondoftwo}%
\providecommand \translation [1]{[#1]}%
\providecommand \BibitemOpen [0]{}%
\providecommand \bibitemStop [0]{}%
\providecommand \bibitemNoStop [0]{.\EOS\space}%
\providecommand \EOS [0]{\spacefactor3000\relax}%
\providecommand \BibitemShut  [1]{\csname bibitem#1\endcsname}%
\let\auto@bib@innerbib\@empty
\bibitem [{\citenamefont {Preskill}(2018)}]{preskillQuantumComputingNISQ2018}%
  \BibitemOpen
  \bibfield  {author} {\bibinfo {author} {\bibfnamefont {J.}~\bibnamefont
  {Preskill}},\ }\href {\doibase 10.22331/q-2018-08-06-79} {\bibfield
  {journal} {\bibinfo  {journal} {Quantum}\ }\textbf {\bibinfo {volume} {2}},\
  \bibinfo {pages} {79} (\bibinfo {year} {2018})},\ \Eprint
  {http://arxiv.org/abs/1801.00862} {arXiv:1801.00862} \BibitemShut {NoStop}%
\bibitem [{\citenamefont {Stokes}\ \emph {et~al.}(2019)\citenamefont {Stokes},
  \citenamefont {Izaac}, \citenamefont {Killoran},\ and\ \citenamefont
  {Carleo}}]{stokesQuantumNaturalGradient2019a}%
  \BibitemOpen
  \bibfield  {author} {\bibinfo {author} {\bibfnamefont {J.}~\bibnamefont
  {Stokes}}, \bibinfo {author} {\bibfnamefont {J.}~\bibnamefont {Izaac}},
  \bibinfo {author} {\bibfnamefont {N.}~\bibnamefont {Killoran}}, \ and\
  \bibinfo {author} {\bibfnamefont {G.}~\bibnamefont {Carleo}},\ }\href@noop {}
  {\bibfield  {journal} {\bibinfo  {journal} {arXiv:1909.02108 [quant-ph,
  stat]}\ } (\bibinfo {year} {2019})},\ \Eprint
  {http://arxiv.org/abs/1909.02108} {arXiv:1909.02108 [quant-ph, stat]}
  \BibitemShut {NoStop}%
\bibitem [{\citenamefont {McClean}\ \emph {et~al.}(2016)\citenamefont
  {McClean}, \citenamefont {Romero}, \citenamefont {Babbush},\ and\
  \citenamefont {{Aspuru-Guzik}}}]{mccleanTheoryVariationalHybrid2016}%
  \BibitemOpen
  \bibfield  {author} {\bibinfo {author} {\bibfnamefont {J.~R.}\ \bibnamefont
  {McClean}}, \bibinfo {author} {\bibfnamefont {J.}~\bibnamefont {Romero}},
  \bibinfo {author} {\bibfnamefont {R.}~\bibnamefont {Babbush}}, \ and\
  \bibinfo {author} {\bibfnamefont {A.}~\bibnamefont {{Aspuru-Guzik}}},\ }\href
  {\doibase 10.1088/1367-2630/18/2/023023} {\bibfield  {journal} {\bibinfo
  {journal} {New Journal of Physics}\ }\textbf {\bibinfo {volume} {18}},\
  \bibinfo {pages} {023023} (\bibinfo {year} {2016})},\ \Eprint
  {http://arxiv.org/abs/1509.04279} {arXiv:1509.04279} \BibitemShut {NoStop}%
\bibitem [{\citenamefont {Peruzzo}\ \emph {et~al.}(2014)\citenamefont
  {Peruzzo}, \citenamefont {McClean}, \citenamefont {Shadbolt}, \citenamefont
  {Yung}, \citenamefont {Zhou}, \citenamefont {Love}, \citenamefont
  {{Aspuru-Guzik}},\ and\ \citenamefont
  {O'Brien}}]{peruzzoVariationalEigenvalueSolver2014}%
  \BibitemOpen
  \bibfield  {author} {\bibinfo {author} {\bibfnamefont {A.}~\bibnamefont
  {Peruzzo}}, \bibinfo {author} {\bibfnamefont {J.}~\bibnamefont {McClean}},
  \bibinfo {author} {\bibfnamefont {P.}~\bibnamefont {Shadbolt}}, \bibinfo
  {author} {\bibfnamefont {M.-H.}\ \bibnamefont {Yung}}, \bibinfo {author}
  {\bibfnamefont {X.-Q.}\ \bibnamefont {Zhou}}, \bibinfo {author}
  {\bibfnamefont {P.~J.}\ \bibnamefont {Love}}, \bibinfo {author}
  {\bibfnamefont {A.}~\bibnamefont {{Aspuru-Guzik}}}, \ and\ \bibinfo {author}
  {\bibfnamefont {J.~L.}\ \bibnamefont {O'Brien}},\ }\href {\doibase
  10.1038/ncomms5213} {\bibfield  {journal} {\bibinfo  {journal} {Nature
  Communications}\ }\textbf {\bibinfo {volume} {5}},\ \bibinfo {pages} {4213}
  (\bibinfo {year} {2014})}\BibitemShut {NoStop}%
\bibitem [{\citenamefont {Farhi}\ \emph {et~al.}(2014)\citenamefont {Farhi},
  \citenamefont {Goldstone},\ and\ \citenamefont
  {Gutmann}}]{farhiQuantumApproximateOptimization2014}%
  \BibitemOpen
  \bibfield  {author} {\bibinfo {author} {\bibfnamefont {E.}~\bibnamefont
  {Farhi}}, \bibinfo {author} {\bibfnamefont {J.}~\bibnamefont {Goldstone}}, \
  and\ \bibinfo {author} {\bibfnamefont {S.}~\bibnamefont {Gutmann}},\
  }\href@noop {} {\bibfield  {journal} {\bibinfo  {journal} {arXiv:1411.4028
  [quant-ph]}\ } (\bibinfo {year} {2014})},\ \Eprint
  {http://arxiv.org/abs/1411.4028} {arXiv:1411.4028 [quant-ph]} \BibitemShut
  {NoStop}%
\bibitem [{\citenamefont {Cao}\ \emph {et~al.}(2019)\citenamefont {Cao},
  \citenamefont {Romero}, \citenamefont {Olson}, \citenamefont {Degroote},
  \citenamefont {Johnson}, \citenamefont {Kieferov{\'a}}, \citenamefont
  {Kivlichan}, \citenamefont {Menke}, \citenamefont {Peropadre}, \citenamefont
  {Sawaya}, \citenamefont {Sim}, \citenamefont {Veis},\ and\ \citenamefont
  {{Aspuru-Guzik}}}]{caoQuantumChemistryAge2019}%
  \BibitemOpen
  \bibfield  {author} {\bibinfo {author} {\bibfnamefont {Y.}~\bibnamefont
  {Cao}}, \bibinfo {author} {\bibfnamefont {J.}~\bibnamefont {Romero}},
  \bibinfo {author} {\bibfnamefont {J.~P.}\ \bibnamefont {Olson}}, \bibinfo
  {author} {\bibfnamefont {M.}~\bibnamefont {Degroote}}, \bibinfo {author}
  {\bibfnamefont {P.~D.}\ \bibnamefont {Johnson}}, \bibinfo {author}
  {\bibfnamefont {M.}~\bibnamefont {Kieferov{\'a}}}, \bibinfo {author}
  {\bibfnamefont {I.~D.}\ \bibnamefont {Kivlichan}}, \bibinfo {author}
  {\bibfnamefont {T.}~\bibnamefont {Menke}}, \bibinfo {author} {\bibfnamefont
  {B.}~\bibnamefont {Peropadre}}, \bibinfo {author} {\bibfnamefont {N.~P.~D.}\
  \bibnamefont {Sawaya}}, \bibinfo {author} {\bibfnamefont {S.}~\bibnamefont
  {Sim}}, \bibinfo {author} {\bibfnamefont {L.}~\bibnamefont {Veis}}, \ and\
  \bibinfo {author} {\bibfnamefont {A.}~\bibnamefont {{Aspuru-Guzik}}},\ }\href
  {\doibase 10.1021/acs.chemrev.8b00803} {\bibfield  {journal} {\bibinfo
  {journal} {Chemical Reviews}\ }\textbf {\bibinfo {volume} {119}},\ \bibinfo
  {pages} {10856} (\bibinfo {year} {2019})}\BibitemShut {NoStop}%
\bibitem [{\citenamefont {Farhi}\ and\ \citenamefont
  {Neven}(2018)}]{farhiClassificationQuantumNeural2018}%
  \BibitemOpen
  \bibfield  {author} {\bibinfo {author} {\bibfnamefont {E.}~\bibnamefont
  {Farhi}}\ and\ \bibinfo {author} {\bibfnamefont {H.}~\bibnamefont {Neven}},\
  }\href@noop {} {\bibfield  {journal} {\bibinfo  {journal} {arXiv:1802.06002
  [quant-ph]}\ } (\bibinfo {year} {2018})},\ \Eprint
  {http://arxiv.org/abs/1802.06002} {arXiv:1802.06002 [quant-ph]} \BibitemShut
  {NoStop}%
\bibitem [{\citenamefont {{P{\'e}rez-Salinas}}\ \emph
  {et~al.}(2020)\citenamefont {{P{\'e}rez-Salinas}}, \citenamefont
  {{Cervera-Lierta}}, \citenamefont {{Gil-Fuster}},\ and\ \citenamefont
  {Latorre}}]{perez-salinasDataReuploadingUniversal2020}%
  \BibitemOpen
  \bibfield  {author} {\bibinfo {author} {\bibfnamefont {A.}~\bibnamefont
  {{P{\'e}rez-Salinas}}}, \bibinfo {author} {\bibfnamefont {A.}~\bibnamefont
  {{Cervera-Lierta}}}, \bibinfo {author} {\bibfnamefont {E.}~\bibnamefont
  {{Gil-Fuster}}}, \ and\ \bibinfo {author} {\bibfnamefont {J.~I.}\
  \bibnamefont {Latorre}},\ }\href {\doibase 10.22331/q-2020-02-06-226}
  {\bibfield  {journal} {\bibinfo  {journal} {Quantum}\ }\textbf {\bibinfo
  {volume} {4}},\ \bibinfo {pages} {226} (\bibinfo {year} {2020})},\ \Eprint
  {http://arxiv.org/abs/1907.02085} {arXiv:1907.02085} \BibitemShut {NoStop}%
\bibitem [{\citenamefont {Tacchino}\ \emph {et~al.}(2019)\citenamefont
  {Tacchino}, \citenamefont {Macchiavello}, \citenamefont {Gerace},\ and\
  \citenamefont {Bajoni}}]{tacchinoArtificialNeuronImplemented2019a}%
  \BibitemOpen
  \bibfield  {author} {\bibinfo {author} {\bibfnamefont {F.}~\bibnamefont
  {Tacchino}}, \bibinfo {author} {\bibfnamefont {C.}~\bibnamefont
  {Macchiavello}}, \bibinfo {author} {\bibfnamefont {D.}~\bibnamefont
  {Gerace}}, \ and\ \bibinfo {author} {\bibfnamefont {D.}~\bibnamefont
  {Bajoni}},\ }\href {\doibase 10.1038/s41534-019-0140-4} {\bibfield  {journal}
  {\bibinfo  {journal} {npj Quantum Information}\ }\textbf {\bibinfo {volume}
  {5}},\ \bibinfo {pages} {26} (\bibinfo {year} {2019})}\BibitemShut {NoStop}%
\bibitem [{\citenamefont {Carleo}\ and\ \citenamefont
  {Troyer}(2017)}]{carleoSolvingQuantumManyBody2017}%
  \BibitemOpen
  \bibfield  {author} {\bibinfo {author} {\bibfnamefont {G.}~\bibnamefont
  {Carleo}}\ and\ \bibinfo {author} {\bibfnamefont {M.}~\bibnamefont
  {Troyer}},\ }\href {\doibase 10.1126/science.aag2302} {\bibfield  {journal}
  {\bibinfo  {journal} {Science}\ }\textbf {\bibinfo {volume} {355}},\ \bibinfo
  {pages} {602} (\bibinfo {year} {2017})},\ \Eprint
  {http://arxiv.org/abs/1606.02318} {arXiv:1606.02318} \BibitemShut {NoStop}%
\bibitem [{\citenamefont {{Hibat-Allah}}\ \emph {et~al.}(2020)\citenamefont
  {{Hibat-Allah}}, \citenamefont {Ganahl}, \citenamefont {Hayward},
  \citenamefont {Melko},\ and\ \citenamefont
  {Carrasquilla}}]{hibat-allahRecurrentNeuralNetwork2020}%
  \BibitemOpen
  \bibfield  {author} {\bibinfo {author} {\bibfnamefont {M.}~\bibnamefont
  {{Hibat-Allah}}}, \bibinfo {author} {\bibfnamefont {M.}~\bibnamefont
  {Ganahl}}, \bibinfo {author} {\bibfnamefont {L.~E.}\ \bibnamefont {Hayward}},
  \bibinfo {author} {\bibfnamefont {R.~G.}\ \bibnamefont {Melko}}, \ and\
  \bibinfo {author} {\bibfnamefont {J.}~\bibnamefont {Carrasquilla}},\
  }\href@noop {} {\bibfield  {journal} {\bibinfo  {journal} {arXiv:2002.02973
  [cond-mat, physics:physics, physics:quant-ph]}\ } (\bibinfo {year} {2020})},\
  \Eprint {http://arxiv.org/abs/2002.02973} {arXiv:2002.02973 [cond-mat,
  physics:physics, physics:quant-ph]} \BibitemShut {NoStop}%
\bibitem [{\citenamefont {Zoufal}\ \emph {et~al.}(2019)\citenamefont {Zoufal},
  \citenamefont {Lucchi},\ and\ \citenamefont
  {Woerner}}]{zoufalQuantumGenerativeAdversarial2019}%
  \BibitemOpen
  \bibfield  {author} {\bibinfo {author} {\bibfnamefont {C.}~\bibnamefont
  {Zoufal}}, \bibinfo {author} {\bibfnamefont {A.}~\bibnamefont {Lucchi}}, \
  and\ \bibinfo {author} {\bibfnamefont {S.}~\bibnamefont {Woerner}},\ }\href
  {\doibase 10.1038/s41534-019-0223-2} {\bibfield  {journal} {\bibinfo
  {journal} {npj Quantum Information}\ }\textbf {\bibinfo {volume} {5}},\
  \bibinfo {pages} {103} (\bibinfo {year} {2019})}\BibitemShut {NoStop}%
\bibitem [{\citenamefont
  {Hopfield}(1982)}]{hopfieldNeuralNetworksPhysical1982}%
  \BibitemOpen
  \bibfield  {author} {\bibinfo {author} {\bibfnamefont {J.~J.}\ \bibnamefont
  {Hopfield}},\ }\href {\doibase 10.1073/pnas.79.8.2554} {\bibfield  {journal}
  {\bibinfo  {journal} {Proceedings of the National Academy of Sciences}\
  }\textbf {\bibinfo {volume} {79}},\ \bibinfo {pages} {2554} (\bibinfo {year}
  {1982})}\BibitemShut {NoStop}%
\bibitem [{\citenamefont {Ballard}\ \emph {et~al.}(2017)\citenamefont
  {Ballard}, \citenamefont {Das}, \citenamefont {Martiniani}, \citenamefont
  {Mehta}, \citenamefont {Sagun}, \citenamefont {Stevenson},\ and\
  \citenamefont {Wales}}]{ballardEnergyLandscapesMachine2017}%
  \BibitemOpen
  \bibfield  {author} {\bibinfo {author} {\bibfnamefont {A.~J.}\ \bibnamefont
  {Ballard}}, \bibinfo {author} {\bibfnamefont {R.}~\bibnamefont {Das}},
  \bibinfo {author} {\bibfnamefont {S.}~\bibnamefont {Martiniani}}, \bibinfo
  {author} {\bibfnamefont {D.}~\bibnamefont {Mehta}}, \bibinfo {author}
  {\bibfnamefont {L.}~\bibnamefont {Sagun}}, \bibinfo {author} {\bibfnamefont
  {J.~D.}\ \bibnamefont {Stevenson}}, \ and\ \bibinfo {author} {\bibfnamefont
  {D.~J.}\ \bibnamefont {Wales}},\ }\href {\doibase 10.1039/C7CP01108C}
  {\bibfield  {journal} {\bibinfo  {journal} {Physical Chemistry Chemical
  Physics}\ }\textbf {\bibinfo {volume} {19}},\ \bibinfo {pages} {12585}
  (\bibinfo {year} {2017})}\BibitemShut {NoStop}%
\bibitem [{\citenamefont {Sagun}\ \emph {et~al.}(2015)\citenamefont {Sagun},
  \citenamefont {Guney}, \citenamefont {Arous},\ and\ \citenamefont
  {LeCun}}]{sagunExplorationsHighDimensional2015}%
  \BibitemOpen
  \bibfield  {author} {\bibinfo {author} {\bibfnamefont {L.}~\bibnamefont
  {Sagun}}, \bibinfo {author} {\bibfnamefont {V.~U.}\ \bibnamefont {Guney}},
  \bibinfo {author} {\bibfnamefont {G.~B.}\ \bibnamefont {Arous}}, \ and\
  \bibinfo {author} {\bibfnamefont {Y.}~\bibnamefont {LeCun}},\ }\href@noop {}
  {\bibfield  {journal} {\bibinfo  {journal} {arXiv:1412.6615 [cs, stat]}\ }
  (\bibinfo {year} {2015})},\ \Eprint {http://arxiv.org/abs/1412.6615}
  {arXiv:1412.6615 [cs, stat]} \BibitemShut {NoStop}%
\bibitem [{\citenamefont {Draxler}\ \emph {et~al.}(2019)\citenamefont
  {Draxler}, \citenamefont {Veschgini}, \citenamefont {Salmhofer},\ and\
  \citenamefont {Hamprecht}}]{draxlerEssentiallyNoBarriers2019}%
  \BibitemOpen
  \bibfield  {author} {\bibinfo {author} {\bibfnamefont {F.}~\bibnamefont
  {Draxler}}, \bibinfo {author} {\bibfnamefont {K.}~\bibnamefont {Veschgini}},
  \bibinfo {author} {\bibfnamefont {M.}~\bibnamefont {Salmhofer}}, \ and\
  \bibinfo {author} {\bibfnamefont {F.~A.}\ \bibnamefont {Hamprecht}},\
  }\href@noop {} {\bibfield  {journal} {\bibinfo  {journal} {arXiv:1803.00885
  [cs, stat]}\ } (\bibinfo {year} {2019})},\ \Eprint
  {http://arxiv.org/abs/1803.00885} {arXiv:1803.00885 [cs, stat]} \BibitemShut
  {NoStop}%
\bibitem [{\citenamefont {Sagun}\ \emph {et~al.}(2018)\citenamefont {Sagun},
  \citenamefont {Evci}, \citenamefont {Guney}, \citenamefont {Dauphin},\ and\
  \citenamefont {Bottou}}]{sagunEmpiricalAnalysisHessian2018}%
  \BibitemOpen
  \bibfield  {author} {\bibinfo {author} {\bibfnamefont {L.}~\bibnamefont
  {Sagun}}, \bibinfo {author} {\bibfnamefont {U.}~\bibnamefont {Evci}},
  \bibinfo {author} {\bibfnamefont {V.~U.}\ \bibnamefont {Guney}}, \bibinfo
  {author} {\bibfnamefont {Y.}~\bibnamefont {Dauphin}}, \ and\ \bibinfo
  {author} {\bibfnamefont {L.}~\bibnamefont {Bottou}},\ }\href@noop {}
  {\bibfield  {journal} {\bibinfo  {journal} {arXiv:1706.04454 [cs]}\ }
  (\bibinfo {year} {2018})},\ \Eprint {http://arxiv.org/abs/1706.04454}
  {arXiv:1706.04454 [cs]} \BibitemShut {NoStop}%
\bibitem [{\citenamefont {Keskar}\ \emph
  {et~al.}(2017{\natexlab{a}})\citenamefont {Keskar}, \citenamefont {Mudigere},
  \citenamefont {Nocedal}, \citenamefont {Smelyanskiy},\ and\ \citenamefont
  {Tang}}]{keskarLargeBatchTrainingDeep2017}%
  \BibitemOpen
  \bibfield  {author} {\bibinfo {author} {\bibfnamefont {N.~S.}\ \bibnamefont
  {Keskar}}, \bibinfo {author} {\bibfnamefont {D.}~\bibnamefont {Mudigere}},
  \bibinfo {author} {\bibfnamefont {J.}~\bibnamefont {Nocedal}}, \bibinfo
  {author} {\bibfnamefont {M.}~\bibnamefont {Smelyanskiy}}, \ and\ \bibinfo
  {author} {\bibfnamefont {P.~T.~P.}\ \bibnamefont {Tang}},\ }\href@noop {}
  {\bibfield  {journal} {\bibinfo  {journal} {arXiv:1609.04836 [cs, math]}\ }
  (\bibinfo {year} {2017}{\natexlab{a}})},\ \Eprint
  {http://arxiv.org/abs/1609.04836} {arXiv:1609.04836 [cs, math]} \BibitemShut
  {NoStop}%
\bibitem [{\citenamefont {Alain}\ \emph {et~al.}(2019)\citenamefont {Alain},
  \citenamefont {Roux},\ and\ \citenamefont
  {Manzagol}}]{alainNegativeEigenvaluesHessian2019}%
  \BibitemOpen
  \bibfield  {author} {\bibinfo {author} {\bibfnamefont {G.}~\bibnamefont
  {Alain}}, \bibinfo {author} {\bibfnamefont {N.~L.}\ \bibnamefont {Roux}}, \
  and\ \bibinfo {author} {\bibfnamefont {P.-A.}\ \bibnamefont {Manzagol}},\
  }\href@noop {} {\bibfield  {journal} {\bibinfo  {journal} {arXiv:1902.02366
  [cs, math, stat]}\ } (\bibinfo {year} {2019})},\ \Eprint
  {http://arxiv.org/abs/1902.02366} {arXiv:1902.02366 [cs, math, stat]}
  \BibitemShut {NoStop}%
\bibitem [{\citenamefont {McClean}\ \emph {et~al.}(2018)\citenamefont
  {McClean}, \citenamefont {Boixo}, \citenamefont {Smelyanskiy}, \citenamefont
  {Babbush},\ and\ \citenamefont {Neven}}]{mccleanBarrenPlateausQuantum2018a}%
  \BibitemOpen
  \bibfield  {author} {\bibinfo {author} {\bibfnamefont {J.~R.}\ \bibnamefont
  {McClean}}, \bibinfo {author} {\bibfnamefont {S.}~\bibnamefont {Boixo}},
  \bibinfo {author} {\bibfnamefont {V.~N.}\ \bibnamefont {Smelyanskiy}},
  \bibinfo {author} {\bibfnamefont {R.}~\bibnamefont {Babbush}}, \ and\
  \bibinfo {author} {\bibfnamefont {H.}~\bibnamefont {Neven}},\ }\href
  {\doibase 10.1038/s41467-018-07090-4} {\bibfield  {journal} {\bibinfo
  {journal} {Nature Communications}\ }\textbf {\bibinfo {volume} {9}},\
  \bibinfo {pages} {4812} (\bibinfo {year} {2018})}\BibitemShut {NoStop}%
\bibitem [{\citenamefont {Cerezo}\ \emph {et~al.}(2020)\citenamefont {Cerezo},
  \citenamefont {Sone}, \citenamefont {Volkoff}, \citenamefont {Cincio},\ and\
  \citenamefont {Coles}}]{cerezoCostFunctionDependentBarrenPlateaus2020}%
  \BibitemOpen
  \bibfield  {author} {\bibinfo {author} {\bibfnamefont {M.}~\bibnamefont
  {Cerezo}}, \bibinfo {author} {\bibfnamefont {A.}~\bibnamefont {Sone}},
  \bibinfo {author} {\bibfnamefont {T.}~\bibnamefont {Volkoff}}, \bibinfo
  {author} {\bibfnamefont {L.}~\bibnamefont {Cincio}}, \ and\ \bibinfo {author}
  {\bibfnamefont {P.~J.}\ \bibnamefont {Coles}},\ }\href@noop {} {\bibfield
  {journal} {\bibinfo  {journal} {arXiv:2001.00550 [quant-ph]}\ } (\bibinfo
  {year} {2020})},\ \Eprint {http://arxiv.org/abs/2001.00550} {arXiv:2001.00550
  [quant-ph]} \BibitemShut {NoStop}%
\bibitem [{\citenamefont {Grant}\ \emph {et~al.}(2019)\citenamefont {Grant},
  \citenamefont {Wossnig}, \citenamefont {Ostaszewski},\ and\ \citenamefont
  {Benedetti}}]{grantInitializationStrategyAddressing2019}%
  \BibitemOpen
  \bibfield  {author} {\bibinfo {author} {\bibfnamefont {E.}~\bibnamefont
  {Grant}}, \bibinfo {author} {\bibfnamefont {L.}~\bibnamefont {Wossnig}},
  \bibinfo {author} {\bibfnamefont {M.}~\bibnamefont {Ostaszewski}}, \ and\
  \bibinfo {author} {\bibfnamefont {M.}~\bibnamefont {Benedetti}},\ }\href
  {\doibase 10.22331/q-2019-12-09-214} {\bibfield  {journal} {\bibinfo
  {journal} {Quantum}\ }\textbf {\bibinfo {volume} {3}},\ \bibinfo {pages}
  {214} (\bibinfo {year} {2019})}\BibitemShut {NoStop}%
\bibitem [{\citenamefont {Wierichs}\ \emph {et~al.}(2020)\citenamefont
  {Wierichs}, \citenamefont {Gogolin},\ and\ \citenamefont
  {Kastoryano}}]{wierichsAvoidingLocalMinima2020b}%
  \BibitemOpen
  \bibfield  {author} {\bibinfo {author} {\bibfnamefont {D.}~\bibnamefont
  {Wierichs}}, \bibinfo {author} {\bibfnamefont {C.}~\bibnamefont {Gogolin}}, \
  and\ \bibinfo {author} {\bibfnamefont {M.}~\bibnamefont {Kastoryano}},\
  }\href@noop {} {\bibfield  {journal} {\bibinfo  {journal} {arXiv:2004.14666
  [quant-ph]}\ } (\bibinfo {year} {2020})},\ \Eprint
  {http://arxiv.org/abs/2004.14666} {arXiv:2004.14666 [quant-ph]} \BibitemShut
  {NoStop}%
\bibitem [{\citenamefont {Mehta}\ \emph {et~al.}(2019)\citenamefont {Mehta},
  \citenamefont {Bukov}, \citenamefont {Wang}, \citenamefont {Day},
  \citenamefont {Richardson}, \citenamefont {Fisher},\ and\ \citenamefont
  {Schwab}}]{mehtaHighbiasLowvarianceIntroduction2019a}%
  \BibitemOpen
  \bibfield  {author} {\bibinfo {author} {\bibfnamefont {P.}~\bibnamefont
  {Mehta}}, \bibinfo {author} {\bibfnamefont {M.}~\bibnamefont {Bukov}},
  \bibinfo {author} {\bibfnamefont {C.-H.}\ \bibnamefont {Wang}}, \bibinfo
  {author} {\bibfnamefont {A.~G.~R.}\ \bibnamefont {Day}}, \bibinfo {author}
  {\bibfnamefont {C.}~\bibnamefont {Richardson}}, \bibinfo {author}
  {\bibfnamefont {C.~K.}\ \bibnamefont {Fisher}}, \ and\ \bibinfo {author}
  {\bibfnamefont {D.~J.}\ \bibnamefont {Schwab}},\ }\href {\doibase
  10.1016/j.physrep.2019.03.001} {\bibfield  {journal} {\bibinfo  {journal}
  {Physics Reports}\ }\textbf {\bibinfo {volume} {810}},\ \bibinfo {pages} {1}
  (\bibinfo {year} {2019})},\ \Eprint {http://arxiv.org/abs/1803.08823}
  {arXiv:1803.08823} \BibitemShut {NoStop}%
\bibitem [{\citenamefont {Le}\ \emph {et~al.}()\citenamefont {Le},
  \citenamefont {Ngiam}, \citenamefont {Coates}, \citenamefont {Lahiri},
  \citenamefont {Prochnow},\ and\ \citenamefont
  {Ng}}]{leOptimizationMethodsDeep}%
  \BibitemOpen
  \bibfield  {author} {\bibinfo {author} {\bibfnamefont {Q.~V.}\ \bibnamefont
  {Le}}, \bibinfo {author} {\bibfnamefont {J.}~\bibnamefont {Ngiam}}, \bibinfo
  {author} {\bibfnamefont {A.}~\bibnamefont {Coates}}, \bibinfo {author}
  {\bibfnamefont {A.}~\bibnamefont {Lahiri}}, \bibinfo {author} {\bibfnamefont
  {B.}~\bibnamefont {Prochnow}}, \ and\ \bibinfo {author} {\bibfnamefont
  {A.~Y.}\ \bibnamefont {Ng}},\ }\href@noop {} {\ ,\ \bibinfo {pages}
  {8}}\BibitemShut {NoStop}%
\bibitem [{\citenamefont {LeCun}\ \emph {et~al.}(1990)\citenamefont {LeCun},
  \citenamefont {Denker},\ and\ \citenamefont {Solla}}]{NIPS1989_250}%
  \BibitemOpen
  \bibfield  {author} {\bibinfo {author} {\bibfnamefont {Y.}~\bibnamefont
  {LeCun}}, \bibinfo {author} {\bibfnamefont {J.~S.}\ \bibnamefont {Denker}}, \
  and\ \bibinfo {author} {\bibfnamefont {S.~A.}\ \bibnamefont {Solla}},\ }in\
  \href {http://papers.nips.cc/paper/250-optimal-brain-damage.pdf} {\emph
  {\bibinfo {booktitle} {Advances in Neural Information Processing Systems}}},\
  \bibinfo {editor} {edited by\ \bibinfo {editor} {\bibfnamefont {D.~S.}\
  \bibnamefont {Touretzky}}}\ (\bibinfo {year} {1990})\ pp.\ \bibinfo {pages}
  {598--605}\BibitemShut {NoStop}%
\bibitem [{\citenamefont {Hassibi}\ and\ \citenamefont
  {Stork}(1993)}]{hassibi1993second}%
  \BibitemOpen
  \bibfield  {author} {\bibinfo {author} {\bibfnamefont {B.}~\bibnamefont
  {Hassibi}}\ and\ \bibinfo {author} {\bibfnamefont {D.~G.}\ \bibnamefont
  {Stork}},\ }in\ \href {\doibase 10.1109/icnn.1993.298572} {\emph {\bibinfo
  {booktitle} {Advances in neural information processing systems}}}\ (\bibinfo
  {year} {1993})\ p.\ \bibinfo {pages} {164}\BibitemShut {NoStop}%
\bibitem [{\citenamefont {Koh}\ and\ \citenamefont
  {Liang}(2017)}]{kohUnderstandingBlackboxPredictions2017b}%
  \BibitemOpen
  \bibfield  {author} {\bibinfo {author} {\bibfnamefont {P.~W.}\ \bibnamefont
  {Koh}}\ and\ \bibinfo {author} {\bibfnamefont {P.}~\bibnamefont {Liang}},\
  }\href@noop {} {\bibfield  {journal} {\bibinfo  {journal} {arXiv:1703.04730
  [cs, stat]}\ } (\bibinfo {year} {2017})},\ \Eprint
  {http://arxiv.org/abs/1703.04730} {arXiv:1703.04730 [cs, stat]} \BibitemShut
  {NoStop}%
\bibitem [{\citenamefont {Sharma}\ \emph {et~al.}(2020)\citenamefont {Sharma},
  \citenamefont {Cerezo}, \citenamefont {Cincio},\ and\ \citenamefont
  {Coles}}]{sharma2020trainability}%
  \BibitemOpen
  \bibfield  {author} {\bibinfo {author} {\bibfnamefont {K.}~\bibnamefont
  {Sharma}}, \bibinfo {author} {\bibfnamefont {M.}~\bibnamefont {Cerezo}},
  \bibinfo {author} {\bibfnamefont {L.}~\bibnamefont {Cincio}}, \ and\ \bibinfo
  {author} {\bibfnamefont {P.~J.}\ \bibnamefont {Coles}},\ }\href@noop {}
  {\bibfield  {journal} {\bibinfo  {journal} {arXiv preprint arXiv:2005.12458}\
  } (\bibinfo {year} {2020})}\BibitemShut {NoStop}%
\bibitem [{\citenamefont {Sim}\ \emph {et~al.}(2019)\citenamefont {Sim},
  \citenamefont {Johnson},\ and\ \citenamefont
  {{Aspuru-Guzik}}}]{simExpressibilityEntanglingCapability2019}%
  \BibitemOpen
  \bibfield  {author} {\bibinfo {author} {\bibfnamefont {S.}~\bibnamefont
  {Sim}}, \bibinfo {author} {\bibfnamefont {P.~D.}\ \bibnamefont {Johnson}}, \
  and\ \bibinfo {author} {\bibfnamefont {A.}~\bibnamefont {{Aspuru-Guzik}}},\
  }\href {\doibase 10.1002/qute.201900070} {\bibfield  {journal} {\bibinfo
  {journal} {Advanced Quantum Technologies}\ }\textbf {\bibinfo {volume} {2}},\
  \bibinfo {pages} {1900070} (\bibinfo {year} {2019})},\ \Eprint
  {http://arxiv.org/abs/1905.10876} {arXiv:1905.10876} \BibitemShut {NoStop}%
\bibitem [{\citenamefont {Keskar}\ \emph
  {et~al.}(2017{\natexlab{b}})\citenamefont {Keskar}, \citenamefont {Mudigere},
  \citenamefont {Nocedal}, \citenamefont {Smelyanskiy},\ and\ \citenamefont
  {Tang}}]{keskarLargeBatchTrainingDeep2017a}%
  \BibitemOpen
  \bibfield  {author} {\bibinfo {author} {\bibfnamefont {N.~S.}\ \bibnamefont
  {Keskar}}, \bibinfo {author} {\bibfnamefont {D.}~\bibnamefont {Mudigere}},
  \bibinfo {author} {\bibfnamefont {J.}~\bibnamefont {Nocedal}}, \bibinfo
  {author} {\bibfnamefont {M.}~\bibnamefont {Smelyanskiy}}, \ and\ \bibinfo
  {author} {\bibfnamefont {P.~T.~P.}\ \bibnamefont {Tang}},\ }\href@noop {}
  {\bibfield  {journal} {\bibinfo  {journal} {arXiv:1609.04836 [cs, math]}\ }
  (\bibinfo {year} {2017}{\natexlab{b}})},\ \Eprint
  {http://arxiv.org/abs/1609.04836} {arXiv:1609.04836 [cs, math]} \BibitemShut
  {NoStop}%
\bibitem [{\citenamefont {Brekelmans}\ \emph {et~al.}(2005)\citenamefont
  {Brekelmans}, \citenamefont {Driessen}, \citenamefont {Hamers},\ and\
  \citenamefont {den. Hertog}}]{brekelmansGradientEstimationSchemes2005}%
  \BibitemOpen
  \bibfield  {author} {\bibinfo {author} {\bibfnamefont {R.~C.~M.}\
  \bibnamefont {Brekelmans}}, \bibinfo {author} {\bibfnamefont {L.~T.}\
  \bibnamefont {Driessen}}, \bibinfo {author} {\bibfnamefont {H.~J.~M.}\
  \bibnamefont {Hamers}}, \ and\ \bibinfo {author} {\bibfnamefont
  {D.}~\bibnamefont {den. Hertog}},\ }\href {\doibase
  10.1007/s10957-005-5496-2} {\bibfield  {journal} {\bibinfo  {journal}
  {Journal of Optimization Theory and Applications}\ }\textbf {\bibinfo
  {volume} {126}},\ \bibinfo {pages} {529} (\bibinfo {year}
  {2005})}\BibitemShut {NoStop}%
\bibitem [{\citenamefont {Mitarai}\ \emph {et~al.}(2018)\citenamefont
  {Mitarai}, \citenamefont {Negoro}, \citenamefont {Kitagawa},\ and\
  \citenamefont {Fujii}}]{mitaraiQuantumCircuitLearning2018}%
  \BibitemOpen
  \bibfield  {author} {\bibinfo {author} {\bibfnamefont {K.}~\bibnamefont
  {Mitarai}}, \bibinfo {author} {\bibfnamefont {M.}~\bibnamefont {Negoro}},
  \bibinfo {author} {\bibfnamefont {M.}~\bibnamefont {Kitagawa}}, \ and\
  \bibinfo {author} {\bibfnamefont {K.}~\bibnamefont {Fujii}},\ }\href
  {\doibase 10.1103/PhysRevA.98.032309} {\bibfield  {journal} {\bibinfo
  {journal} {Physical Review A}\ }\textbf {\bibinfo {volume} {98}},\ \bibinfo
  {pages} {032309} (\bibinfo {year} {2018})},\ \Eprint
  {http://arxiv.org/abs/1803.00745} {arXiv:1803.00745} \BibitemShut {NoStop}%
\bibitem [{\citenamefont {Schuld}\ \emph {et~al.}(2019)\citenamefont {Schuld},
  \citenamefont {Bergholm}, \citenamefont {Gogolin}, \citenamefont {Izaac},\
  and\ \citenamefont {Killoran}}]{schuldEvaluatingAnalyticGradients2019}%
  \BibitemOpen
  \bibfield  {author} {\bibinfo {author} {\bibfnamefont {M.}~\bibnamefont
  {Schuld}}, \bibinfo {author} {\bibfnamefont {V.}~\bibnamefont {Bergholm}},
  \bibinfo {author} {\bibfnamefont {C.}~\bibnamefont {Gogolin}}, \bibinfo
  {author} {\bibfnamefont {J.}~\bibnamefont {Izaac}}, \ and\ \bibinfo {author}
  {\bibfnamefont {N.}~\bibnamefont {Killoran}},\ }\href {\doibase
  10.1103/PhysRevA.99.032331} {\bibfield  {journal} {\bibinfo  {journal}
  {Physical Review A}\ }\textbf {\bibinfo {volume} {99}},\ \bibinfo {pages}
  {032331} (\bibinfo {year} {2019})},\ \Eprint
  {http://arxiv.org/abs/1811.11184} {arXiv:1811.11184} \BibitemShut {NoStop}%
\bibitem [{\citenamefont {Mitarai}\ and\ \citenamefont
  {Fujii}(2019)}]{mitarai2019methodology}%
  \BibitemOpen
  \bibfield  {author} {\bibinfo {author} {\bibfnamefont {K.}~\bibnamefont
  {Mitarai}}\ and\ \bibinfo {author} {\bibfnamefont {K.}~\bibnamefont
  {Fujii}},\ }\href@noop {} {\bibfield  {journal} {\bibinfo  {journal}
  {Physical Review Research}\ }\textbf {\bibinfo {volume} {1}},\ \bibinfo
  {pages} {013006} (\bibinfo {year} {2019})}\BibitemShut {NoStop}%
\bibitem [{\citenamefont {Sagun}\ \emph {et~al.}(2017)\citenamefont {Sagun},
  \citenamefont {Bottou},\ and\ \citenamefont
  {LeCun}}]{sagunEigenvaluesHessianDeep2017}%
  \BibitemOpen
  \bibfield  {author} {\bibinfo {author} {\bibfnamefont {L.}~\bibnamefont
  {Sagun}}, \bibinfo {author} {\bibfnamefont {L.}~\bibnamefont {Bottou}}, \
  and\ \bibinfo {author} {\bibfnamefont {Y.}~\bibnamefont {LeCun}},\
  }\href@noop {} {\bibfield  {journal} {\bibinfo  {journal} {arXiv:1611.07476
  [cs]}\ } (\bibinfo {year} {2017})},\ \Eprint
  {http://arxiv.org/abs/1611.07476} {arXiv:1611.07476 [cs]} \BibitemShut
  {NoStop}%
\bibitem [{\citenamefont {Chaudhari}\ \emph {et~al.}(2017)\citenamefont
  {Chaudhari}, \citenamefont {Choromanska}, \citenamefont {Soatto},
  \citenamefont {LeCun}, \citenamefont {Baldassi}, \citenamefont {Borgs},
  \citenamefont {Chayes}, \citenamefont {Sagun},\ and\ \citenamefont
  {Zecchina}}]{chaudhariEntropySGDBiasingGradient2017}%
  \BibitemOpen
  \bibfield  {author} {\bibinfo {author} {\bibfnamefont {P.}~\bibnamefont
  {Chaudhari}}, \bibinfo {author} {\bibfnamefont {A.}~\bibnamefont
  {Choromanska}}, \bibinfo {author} {\bibfnamefont {S.}~\bibnamefont {Soatto}},
  \bibinfo {author} {\bibfnamefont {Y.}~\bibnamefont {LeCun}}, \bibinfo
  {author} {\bibfnamefont {C.}~\bibnamefont {Baldassi}}, \bibinfo {author}
  {\bibfnamefont {C.}~\bibnamefont {Borgs}}, \bibinfo {author} {\bibfnamefont
  {J.}~\bibnamefont {Chayes}}, \bibinfo {author} {\bibfnamefont
  {L.}~\bibnamefont {Sagun}}, \ and\ \bibinfo {author} {\bibfnamefont
  {R.}~\bibnamefont {Zecchina}},\ }\href@noop {} {\bibfield  {journal}
  {\bibinfo  {journal} {arXiv:1611.01838 [cs, stat]}\ } (\bibinfo {year}
  {2017})},\ \Eprint {http://arxiv.org/abs/1611.01838} {arXiv:1611.01838 [cs,
  stat]} \BibitemShut {NoStop}%
\bibitem [{\citenamefont {Li}\ \emph {et~al.}(2020)\citenamefont {Li},
  \citenamefont {Ding},\ and\ \citenamefont {Sun}}]{liBenefitWidthNeural2020}%
  \BibitemOpen
  \bibfield  {author} {\bibinfo {author} {\bibfnamefont {D.}~\bibnamefont
  {Li}}, \bibinfo {author} {\bibfnamefont {T.}~\bibnamefont {Ding}}, \ and\
  \bibinfo {author} {\bibfnamefont {R.}~\bibnamefont {Sun}},\ }\href@noop {}
  {\bibfield  {journal} {\bibinfo  {journal} {arXiv:1812.11039 [cs, math,
  stat]}\ } (\bibinfo {year} {2020})},\ \Eprint
  {http://arxiv.org/abs/1812.11039} {arXiv:1812.11039 [cs, math, stat]}
  \BibitemShut {NoStop}%
\bibitem [{\citenamefont {Schuld}\ \emph {et~al.}(2020)\citenamefont {Schuld},
  \citenamefont {Sweke},\ and\ \citenamefont {Meyer}}]{schuld2020effect}%
  \BibitemOpen
  \bibfield  {author} {\bibinfo {author} {\bibfnamefont {M.}~\bibnamefont
  {Schuld}}, \bibinfo {author} {\bibfnamefont {R.}~\bibnamefont {Sweke}}, \
  and\ \bibinfo {author} {\bibfnamefont {J.~J.}\ \bibnamefont {Meyer}},\
  }\href@noop {} {\bibfield  {journal} {\bibinfo  {journal} {arXiv preprint
  arXiv:2008.08605}\ } (\bibinfo {year} {2020})}\BibitemShut {NoStop}%
\bibitem [{\citenamefont {Bergholm}\ \emph {et~al.}(2020)\citenamefont
  {Bergholm}, \citenamefont {Izaac}, \citenamefont {Schuld}, \citenamefont
  {Gogolin}, \citenamefont {Alam}, \citenamefont {Ahmed}, \citenamefont
  {Arrazola}, \citenamefont {Blank}, \citenamefont {Delgado}, \citenamefont
  {Jahangiri}, \citenamefont {McKiernan}, \citenamefont {Meyer}, \citenamefont
  {Niu}, \citenamefont {Sz{\'a}va},\ and\ \citenamefont
  {Killoran}}]{bergholmPennyLaneAutomaticDifferentiation2020}%
  \BibitemOpen
  \bibfield  {author} {\bibinfo {author} {\bibfnamefont {V.}~\bibnamefont
  {Bergholm}}, \bibinfo {author} {\bibfnamefont {J.}~\bibnamefont {Izaac}},
  \bibinfo {author} {\bibfnamefont {M.}~\bibnamefont {Schuld}}, \bibinfo
  {author} {\bibfnamefont {C.}~\bibnamefont {Gogolin}}, \bibinfo {author}
  {\bibfnamefont {M.~S.}\ \bibnamefont {Alam}}, \bibinfo {author}
  {\bibfnamefont {S.}~\bibnamefont {Ahmed}}, \bibinfo {author} {\bibfnamefont
  {J.~M.}\ \bibnamefont {Arrazola}}, \bibinfo {author} {\bibfnamefont
  {C.}~\bibnamefont {Blank}}, \bibinfo {author} {\bibfnamefont
  {A.}~\bibnamefont {Delgado}}, \bibinfo {author} {\bibfnamefont
  {S.}~\bibnamefont {Jahangiri}}, \bibinfo {author} {\bibfnamefont
  {K.}~\bibnamefont {McKiernan}}, \bibinfo {author} {\bibfnamefont {J.~J.}\
  \bibnamefont {Meyer}}, \bibinfo {author} {\bibfnamefont {Z.}~\bibnamefont
  {Niu}}, \bibinfo {author} {\bibfnamefont {A.}~\bibnamefont {Sz{\'a}va}}, \
  and\ \bibinfo {author} {\bibfnamefont {N.}~\bibnamefont {Killoran}},\
  }\href@noop {} {\bibfield  {journal} {\bibinfo  {journal} {arXiv:1811.04968
  [physics, physics:quant-ph]}\ } (\bibinfo {year} {2020})},\ \Eprint
  {http://arxiv.org/abs/1811.04968} {arXiv:1811.04968 [physics,
  physics:quant-ph]} \BibitemShut {NoStop}%
\bibitem [{\citenamefont {Beach}\ \emph {et~al.}(2019)\citenamefont {Beach},
  \citenamefont {De~Vlugt}, \citenamefont {Golubeva}, \citenamefont {Huembeli},
  \citenamefont {Kulchytskyy}, \citenamefont {Luo}, \citenamefont {Melko},
  \citenamefont {Merali},\ and\ \citenamefont
  {Torlai}}]{beachQuCumberWavefunctionReconstruction2019}%
  \BibitemOpen
  \bibfield  {author} {\bibinfo {author} {\bibfnamefont {M.~J.~S.}\
  \bibnamefont {Beach}}, \bibinfo {author} {\bibfnamefont {I.}~\bibnamefont
  {De~Vlugt}}, \bibinfo {author} {\bibfnamefont {A.}~\bibnamefont {Golubeva}},
  \bibinfo {author} {\bibfnamefont {P.}~\bibnamefont {Huembeli}}, \bibinfo
  {author} {\bibfnamefont {B.}~\bibnamefont {Kulchytskyy}}, \bibinfo {author}
  {\bibfnamefont {X.}~\bibnamefont {Luo}}, \bibinfo {author} {\bibfnamefont
  {R.}~\bibnamefont {Melko}}, \bibinfo {author} {\bibfnamefont
  {E.}~\bibnamefont {Merali}}, \ and\ \bibinfo {author} {\bibfnamefont
  {G.}~\bibnamefont {Torlai}},\ }\href {\doibase 10.21468/SciPostPhys.7.1.009}
  {\bibfield  {journal} {\bibinfo  {journal} {SciPost Physics}\ }\textbf
  {\bibinfo {volume} {7}},\ \bibinfo {pages} {009} (\bibinfo {year}
  {2019})}\BibitemShut {NoStop}%
\bibitem [{\citenamefont {Huembeli}\ and\ \citenamefont
  {Dauphin}(2020)}]{Huembeli_Github_2020}%
  \BibitemOpen
  \bibfield  {author} {\bibinfo {author} {\bibfnamefont {P.}~\bibnamefont
  {Huembeli}}\ and\ \bibinfo {author} {\bibfnamefont {A.}~\bibnamefont
  {Dauphin}},\ }\href@noop {} {\enquote {\bibinfo {title}
  {{PatrickHuembeli/vqc\_loss\_landscapes: ArXiv\_version\_v1.1}},}\ }\bibinfo
  {howpublished} {\url{https://github.com/PatrickHuembeli/vqc_loss_landscapes}}
  (\bibinfo {year} {2020})\BibitemShut {NoStop}%
\bibitem [{\citenamefont {Jastrzebski}\ \emph {et~al.}(2020)\citenamefont
  {Jastrzebski}, \citenamefont {Szymczak}, \citenamefont {Fort}, \citenamefont
  {Arpit}, \citenamefont {Tabor}, \citenamefont {Cho},\ and\ \citenamefont
  {Geras}}]{jastrzebskiBreakEvenPointOptimization2020}%
  \BibitemOpen
  \bibfield  {author} {\bibinfo {author} {\bibfnamefont {S.}~\bibnamefont
  {Jastrzebski}}, \bibinfo {author} {\bibfnamefont {M.}~\bibnamefont
  {Szymczak}}, \bibinfo {author} {\bibfnamefont {S.}~\bibnamefont {Fort}},
  \bibinfo {author} {\bibfnamefont {D.}~\bibnamefont {Arpit}}, \bibinfo
  {author} {\bibfnamefont {J.}~\bibnamefont {Tabor}}, \bibinfo {author}
  {\bibfnamefont {K.}~\bibnamefont {Cho}}, \ and\ \bibinfo {author}
  {\bibfnamefont {K.}~\bibnamefont {Geras}},\ }\href@noop {} {\bibfield
  {journal} {\bibinfo  {journal} {arXiv:2002.09572 [cs, stat]}\ } (\bibinfo
  {year} {2020})},\ \Eprint {http://arxiv.org/abs/2002.09572} {arXiv:2002.09572
  [cs, stat]} \BibitemShut {NoStop}%
\bibitem [{\citenamefont {Dawid}\ \emph {et~al.}(2020)\citenamefont {Dawid},
  \citenamefont {Huembeli}, \citenamefont {Tomza}, \citenamefont {Lewenstein},\
  and\ \citenamefont {Dauphin}}]{dawidPhaseDetectionNeural2020}%
  \BibitemOpen
  \bibfield  {author} {\bibinfo {author} {\bibfnamefont {A.}~\bibnamefont
  {Dawid}}, \bibinfo {author} {\bibfnamefont {P.}~\bibnamefont {Huembeli}},
  \bibinfo {author} {\bibfnamefont {M.}~\bibnamefont {Tomza}}, \bibinfo
  {author} {\bibfnamefont {M.}~\bibnamefont {Lewenstein}}, \ and\ \bibinfo
  {author} {\bibfnamefont {A.}~\bibnamefont {Dauphin}},\ }\href@noop {}
  {\bibfield  {journal} {\bibinfo  {journal} {arXiv:2004.04711 [cond-mat,
  physics:quant-ph]}\ } (\bibinfo {year} {2020})},\ \Eprint
  {http://arxiv.org/abs/2004.04711} {arXiv:2004.04711 [cond-mat,
  physics:quant-ph]} \BibitemShut {NoStop}%
\bibitem [{\citenamefont {Park}\ and\ \citenamefont
  {Kastoryano}(2020)}]{Park_2020}%
  \BibitemOpen
  \bibfield  {author} {\bibinfo {author} {\bibfnamefont {C.-Y.}\ \bibnamefont
  {Park}}\ and\ \bibinfo {author} {\bibfnamefont {M.~J.}\ \bibnamefont
  {Kastoryano}},\ }\href {\doibase 10.1103/PhysRevResearch.2.023232} {\bibfield
   {journal} {\bibinfo  {journal} {Phys. Rev. Research}\ }\textbf {\bibinfo
  {volume} {2}},\ \bibinfo {pages} {023232} (\bibinfo {year}
  {2020})}\BibitemShut {NoStop}%
\bibitem [{\citenamefont {Cerezo}\ and\ \citenamefont
  {Coles}(2020)}]{cerezoImpactBarrenPlateaus2020}%
  \BibitemOpen
  \bibfield  {author} {\bibinfo {author} {\bibfnamefont {M.}~\bibnamefont
  {Cerezo}}\ and\ \bibinfo {author} {\bibfnamefont {P.~J.}\ \bibnamefont
  {Coles}},\ }\href@noop {} {\bibfield  {journal} {\bibinfo  {journal}
  {arXiv:2008.07454 [quant-ph]}\ } (\bibinfo {year} {2020})},\ \Eprint
  {http://arxiv.org/abs/2008.07454} {arXiv:2008.07454 [quant-ph]} \BibitemShut
  {NoStop}%
\bibitem [{\citenamefont {Mari}\ \emph {et~al.}(2020)\citenamefont {Mari},
  \citenamefont {Bromley},\ and\ \citenamefont
  {Killoran}}]{mariEstimatingGradientHigherorder2020}%
  \BibitemOpen
  \bibfield  {author} {\bibinfo {author} {\bibfnamefont {A.}~\bibnamefont
  {Mari}}, \bibinfo {author} {\bibfnamefont {T.~R.}\ \bibnamefont {Bromley}}, \
  and\ \bibinfo {author} {\bibfnamefont {N.}~\bibnamefont {Killoran}},\
  }\href@noop {} {\bibfield  {journal} {\bibinfo  {journal} {arXiv:2008.06517
  [quant-ph]}\ } (\bibinfo {year} {2020})},\ \Eprint
  {http://arxiv.org/abs/2008.06517} {arXiv:2008.06517 [quant-ph]} \BibitemShut
  {NoStop}%
\end{thebibliography}%

\end{document}